\documentclass[12pt]{article}

\usepackage{graphicx}
\usepackage{amsmath,amssymb}

\makeatletter \@addtoreset{equation}{section}

\renewcommand{\thefootnote}{\fnsymbol{footnote}}
\newcommand{\be}{\begin{equation}}
\newcommand{\ee}{\end{equation}}
\newcommand{\bear}{\begin{eqnarray}}
\newcommand{\eear}{\end{eqnarray}}
\newcommand{\ba}{\begin{array}}
\newcommand{\ea}{\end{array}}

\newcommand{\tr}{{\rm tr}}

\newcommand{\bsubeq}{\begin{subequations}}
\newcommand{\esubeq}{\end{subequations}}

\def\tr{{\rm tr}}

\def\calA{\cal A}
\def\prl{Phys. Rev. Lett.}

\def\lsim{\mathrel{\rlap{\lower3pt\hbox{\hskip1pt$\sim$}}
     \raise1pt\hbox{$<$}}} %less than or approx. symbol
\def\gsim{\mathrel{\rlap{\lower3pt\hbox{\hskip1pt$\sim$}}
     \raise1pt\hbox{$>$}}}
\def\HLS1{HLS$_1$}

\DeclareFontFamily{U}{rsf}{}
\DeclareFontShape{U}{rsf}{m}{n}{
  <5> <6> rsfs5 <7> <8> <9> rsfs7 <10-> rsfs10}{}
\DeclareMathAlphabet\Scr{U}{rsf}{m}{n}
\begin{document}
\begin{titlepage}
\vfill
\begin{flushright}
{\tt\normalsize KIAS-P08030}\\
{\tt\normalsize SU-ITP-08/03}\\
\end{flushright}
\vfill
\begin{center}
{\Large\bf  A Holographic QCD and \\
Excited Baryons from String Theory}

\vfill

Jaemo Park$^{\heartsuit\diamondsuit\spadesuit}$\footnote{\tt jaemo@postech.ac.kr}
and Piljin Yi$^{\clubsuit\spadesuit}$\footnote{\tt piljin@kias.re.kr}

\vskip 5mm
$^{\heartsuit}${\it Department of Physics, Postech, Pohang 790-784,
  Korea }
\vskip 2mm
$^\diamondsuit${\it Postech Center for Theoretical Physics (PCTP), Postech, Pohang
  790-784, Korea}
\vskip 2mm
$^\clubsuit${\it  School of Physics, Korea Institute for Advanced Study, Seoul 130-722, Korea}
\vskip 2mm
$^{\spadesuit}${\it  Department of Physics, Stanford University, Stanford, CA 94305-4060, USA}

\end{center}
\vfill

\begin{abstract}
\noindent We study baryons of arbitrary isospin in a stringy
holographic QCD model. In this D4-D8 holographic setting, the
flavor symmetry is promoted to a gauge symmetry in the bulk, and
produces, as KK modes of the gauge field, pions and spin one
mesons of low energy QCD. Baryons of arbitrary isospins are
represented as instanton solitons with isospin and spin quantum
numbers locked, in a manner similar to the Skyrmion model. The
soliton picture  leads to a natural effective field theory of arbitrary
baryons interacting with mesons. Couplings of baryons to axial
mesons, including pions, are dominated in the large $N_c$ limit by
a direct coupling to the flavor field strength in five dimensions. We delineate
the relevant couplings and determine their strengths. This work
generalizes part of Refs.~\cite{Hong:2007kx,Hong:2007ay} to all
excited baryons. Due to technical difficulties in introducing
relativistic higher spin fields, we perform all computations
in the nonrelativistic regime, which suffices for the leading
$N_c$ predictions.

\end{abstract}

\vfill
\end{titlepage}

\tableofcontents
\renewcommand{\thefootnote}{\#\arabic{footnote}}
\setcounter{footnote}{0}

\section{A Holographic QCD}

The model starts with a stack of $N_c$ D4 branes compactified on a thermal circle.
Because the fermions are given anti-periodic boundary condition, the
massless part of the theory is pure $U(N_c)$ Yang-Mills theory. Scalars
would be also massless classically but due to  the broken supersymmetry
they would acquire mass perturbatively, whereas the gauge fields remain
massless, protected by the gauge symmetry.

Let us first introduce notations for various spacetime coordinates
and indices. The Minkowskii coordinates in which the QCD lives and
in which the noncompact part of $N_c$ D4 branes lives, will be
denoted as
$$
x^\mu,\quad \mu=0,1,2,3 \;,
$$
while the spatial coordinates will be labelled as
$$x^i,\quad i=1,2,3 \qquad \hbox{or}\:\quad x^a, \quad a=1,2,3 \,.$$
We will be forced to mix $a,b,c$ indices, usually reserved as $SU(2)$
gauge indices, and the spatial $i,j,k$ due to the spin-isospin mixing
of the baryon.
The holographic direction provides another spatial direction, whose
coordinate will be either $U$ or $w$. $w$ is the particular choice,
where the relevant five-dimensional geometry has a conformally flat
coordinate $(x^\mu, w)$. Adding this fifth coordinate, we have
$$x^{\hat M},\quad \hat M=0,1,2,3,4\qquad \hbox{or}\quad x^M,\quad M=0,1,2,3,4 \;,$$
and
$$x^m,\quad m=1,2,3,4\;,$$ where $x^4=w$.
The hatted indices, as in $x^{\hat M}$, are raised and lowered using
the proper induced (conformally flat) metric on the D-brane, whereas
unhatted indices are raised and
lowered using the flat metric. The rest of the stringy ten dimensions are
spanned by $S^4$ and one angle, $\tau$, which is the thermal circle wrapped
by the $N_c$ D4 branes.

In the large $N_c$ limit, the dynamics of these D4 is dual to a closed
string theory in some
curved background with flux in accordance with the general AdS/CFT
idea \cite{Maldacena:1997re}.
In the large 't~Hooft coupling limit, $\lambda\equiv g_{YM}^2N_c\gg1$,
and neglecting the gravitational back-reaction from the D8 branes, the
metric is \cite{Witten:1998zw}
\begin{equation}
ds^2=\left(\frac{U}{R}\right)^{3/2}\left(\eta_{\mu\nu}dx^{\mu}dx^{\nu}+f(U)d\tau^2\right)
+\left(\frac{R}{U}\right)^{3/2}\left(\frac{dU^2}{f(U)}+U^2d\Omega_4^2\right) \;,
\end{equation}
with $R^3=\pi g_sN_cl_s^3$ and $f(U)=1-(U_{KK}/U)^3$. The
coordinate $\tau$ is compactified as $\tau=\tau+\delta\tau$ with
$\delta\tau=4\pi R^{3/2}/(3U_{KK}^{1/2})$. The lowest energy sector
of this dual geometry encodes low energy theory of pure $SU(N_c)$
Yang-Mills theory. Glueball spectrum from this dual setup has been
computed with some successful predictions against lattice results
\cite{Csaki:1998qr,Brower:2000rp}.

To add mesons, we introduce
the $N_F$ D8 branes sharing the coordinates $x^0,x^1,x^2,x^3$
with the D4 branes\cite{sakai-sugimoto}.
This allows massless
quark degrees of freedom as open strings attached  to both the D4
and D8 branes. As the D4's are replaced by the geometry, however,
the 4-8 open strings are paired into 8-8 open strings, which are
naturally identified as bi-quark mesons. From the viewpoint of D8 branes,
these mesons arise out of a $U(N_F)$ Yang-Mills theory with the extra
Chern-Simons coupling,
\begin{equation}
\mu_8\int\,\sum C_{p+1} \wedge {\rm Tr}\,e^{2\pi\alpha' {\cal F} }\,,
\end{equation}
on the D8 branes.
We defined $\mu_p={2\pi}/{(2\pi l_s)^{p+1}} $, and $l_s^2=\alpha'$.
$C_{p+1}$'s are the antisymmetric Ramond-Ramond
fields.

The induced metric on the D8 brane is
\begin{equation}
g_{8+1}=\left(\frac{U}{R}\right)^{3/2}\left(\eta_{\mu\nu}dx^{\mu}dx^{\nu}\right)
+\left(\frac{R}{U}\right)^{3/2}\left(\frac{dU^2}{f(U)}+U^2d\Omega_4^2\right) \: .
\end{equation}
A useful choice of the coordinate is \footnote{Another choice of the radial coordinate
$z$ defined as
\begin{equation}
U^3=U_{KK}^3+U_{KK}z^2 \:  ,
\end{equation}
was used by Sakai and Sugimoto\cite{sakai-sugimoto}.
Near origin $w\simeq 0$, we have
$M_{KK}w\simeq {z}/{U_{KK}}$.}
%$H(w)=K(z)^{-1/6}$ with
\begin{equation}
w=\int_{U_{KK}}^U\frac{R^{3/2}dU^\prime}{\sqrt{{U^\prime}^3-U_{KK}^3}}\:.
\end{equation}
with which we have
\begin{equation}
g_{8+1}=\frac{U^{3/2}(w)}{R^{3/2}}\left(dw^2+\eta_{\mu\nu}dx^{\mu}dx^{\nu}\right)
+\frac{R^{3/2}}{U^{1/2}(w)} d\Omega_4^2\:.
\end{equation}
The noncompact part of the D8 brane worldvolume is conformally
equivalent to an interval $[-w_{max}, w_{max}]$ times $R^{3+1}$ with
\begin{equation}
w_{max}=\int_0^\infty\frac{R^{3/2}dU}{\sqrt{U^3-U_{KK}^3}}
=\frac{1}{M_{KK}}\frac32\int_1^\infty\frac{d\tilde U}{\sqrt{\tilde U^3-1}}
\end{equation}
which makes the search for exact instanton solution rather
problematic.

Let us list parameters of the background.
We have
\begin{equation}
R^3=\frac{g_{YM}^2N_cl_s^2}{2M_{KK}}\:,\qquad U_{KK}=\frac{2g_{YM}^2N_cM_{KK}l_s^2}{9}\: ,
\end{equation}
so that
$
M_{KK}\equiv 3U_{KK}^{1/2}/2R^{3/2} \:.
$
Also the nominal Yang-Mills coupling $g_{YM}^2$ is related to the other
parameters as
\begin{equation}
g_{YM}^2=2\pi g_s M_{KK} l_s \:.\,
\end{equation}
where $g_s$ is the string coupling,
but is not a physical parameter on its own. The low energy parameters
of this holographic theory are $M_{KK}$ and $\lambda$, which together
with $N_c$ sets all the physical scales such as the QCD scale and the
pion decay constant. Another important quantity to have in mind is
\begin{equation}
l_s^{warped}\equiv l_s\times (R/U_{KK})^{3/4}\simeq \frac{2.6}{M_{KK}\sqrt{\lambda}}\:,
\end{equation}
which is basically the warped string length scale. This is the string
length scale as measured by $x^\mu$ coordinates at $U=U_{KK}$.

In the low energy limit, the worldvolume dynamics of the
D8 brane is described in terms of a derivative expansion
of the full stringy effective action.
The effective action is
\begin{eqnarray}
-\frac14\;\int_{4+1}\sqrt{-g_{4+1}}
\;\frac{e^{-\Phi}V_{S^4}}{2\pi (2\pi l_s)^5} \;\tr {\cal F}_{\hat M \hat
N}{\cal F}^{\hat M\hat N} + \frac{N_c}{24\pi^2}\int_{4+1}\omega_{5}({\cal A})\: ,\label{dbi}
\end{eqnarray}
with $d\omega_5({\cal A})=\tr {\cal F}^3$.
Here $V_{S^4}$ is the position-dependent volume of the compact
part with
\begin{equation}
V_{S^4}=\frac{8\pi^2}{3}R^3U \: ,
\end{equation}
while the dilaton is
\begin{equation}
e^{-\Phi}=\frac{1}{g_s}\left(\frac{R}{U}\right)^{3/4} \:.
\end{equation}
The Chern-Simons coupling arises from the second set of terms
because $\int_{S^4}dC_3\sim N_c$ takes a quantized value, and was
worked out by Sakai and Sugimoto in some detail \cite{sakai-sugimoto}.

The massless sector  upon dimensional reduction to
four dimension produces the Chiral
lagrangian with a Skyrme term\cite{skyrme}. The pion field $\pi$ is conveniently
expressed in the exponentiated forms
\begin{equation}
U(x)=e^{2i\pi(x) /f_\pi} \: ,\xi(x)=e^{i\pi(x) /f_\pi}\:,
\end{equation}
which can be found in the five-dimensional gauge field
in the following expansion, with the gauge choice ${\cal A}_w=0$,
\begin{equation}
{\cal A}_\mu(x;w)= i\alpha_\mu(x)\psi_0(w)+i\beta_\mu(x) +\sum_n
a_\mu^{(n)}(x)\psi_{(n)}(w)\:,
\end{equation}
where the $SU(N_F)$ part of the lowest lying modes are directly
connected to the pion field as
\begin{equation}
 \alpha_\mu(x)^{SU(N_F)}\equiv \{\xi^{-1},\partial_\mu\xi\}\simeq {2i\over
f_\pi}\partial_\mu\pi ,\qquad
\beta_\mu(x)^{SU(N_F)}\equiv\frac{1}{2}[\xi^{-1},\partial_\mu\xi]\simeq
{1\over 2f_{\pi}^2} [\pi, \partial_\mu\pi]\:,
\end{equation}
where $\psi_0(w)=\psi_0(w(z))={1\over\pi}\arctan\left(z\over
U_{KK}\right)$.
Truncating to pions only, this reproduces the Skyrme Lagrangian\cite{skyrme}
\begin{equation}
{\cal L}_{pion}={f_\pi^2\over 4}\tr \left(U^{-1}\partial_\mu
U\right)^2 +{1\over 32 e^2_{Skyr
me}} \tr \left[ U^{-1}\partial_\mu U,
U^{-1} \partial_\nu U \right]^2\:,
\end{equation}
with
\begin{equation}
f_\pi^2={1\over 54\pi^4}(g_{YM}^2 N_c) N_c M_{KK}^2\:,\quad
e^2_{Skyrme}\simeq  {54\pi^7\over 61} {1\over (g_{YM}^2 N_c)N_c}\:.
\end{equation}
For the rest of KK tower, which are vector mesons and axial vector mesons,
we have the standard kinetic term
\begin{equation}
{\cal L}_{vectors}^{free}=\sum_n\left\{{1\over 4} {\cal F}_{\mu\nu}^{(n)}
{\cal F}^{\mu\nu(n)}+{1\over 2}m_n^2 a_\mu^{(n)} a^{\mu(n)}\right\}\:,
\end{equation}
with ${\cal F}^{(n)}_{\mu\nu}=\partial_\mu a^{(n)}_\nu-\partial_\nu
a^{(n)}_\mu$, plus various interactions between them as well as
with pions.
 Finally there is the WZW term ${\cal L}_{WZW}$ also,
arising from the Chern-Simons term, details of which
can be found in \cite{sakai-sugimoto}.

One should in principle include other hadrons of QCD to this
picture to have a complete QCD-like theory. Glueballs are already
present in the setup as the gravity part of the holographic theory
but a systematic study of glueball/meson interaction is not available
beyond the initial but interesting study in Ref.~\cite{Hashimoto:2007ze}.
On the other hand, experimentally, proper identification of glueballs
is not available so comparison with data is not easy. The other, obvious,
set of hadrons are baryons whose properties have been explored recently
\cite{Hong:2007kx,Hata:2007mb,Hong:2007ay}. In this work, we wish to
generalize and expand these recent studies to baryons of arbitrary
isospins.\footnote{ Another interesting direction of study involving
holographic baryon in Sakai-Sugimoto model is to consider physics in
the background of a baryon density. See for example recent works in
Ref.~\cite{Kim:2008zn,Bergman:2008sg,Kim:2007zm}.}

\section{A Note on the Effective Field Theory Approach}

We reviewed above how the low energy effective theory of mesons
emerges from this holographic setup. Before we go into the discussion of
baryons, it is worthwhile to clarify how we are meant to
use the effective action thus derived. The AdS/CFT
correspondence in general is meant to be a conjectured duality between
an open string theory and a closed string theory. As such,
we anticipate such a correspondence at full quantum level on
both sides. In practice, however, we often must resort to large $N_c$
and large 't Hooft coupling limit where we at least can compute
quantities on the closed string side. This usual limit allows us
to treat the closed string side as a classical theory of gravity
and its multiplets \cite{Maldacena:1997re}.

Approaches to holographic QCD so far have not escaped this limitation.
As a result, when we consider the same large $N_c$ and large $\lambda$
limit, the so-called "bulk side" is meant to be used classically.
The effective field theory such as above is derived strictly in
this spirit, and is meant to be used classically. In
other words, we should not try to renormalize it further by
computing loop diagrams. We are only allowed to compute tree-level
amplitudes using such the vertices present in the effective action. In this sense, the
effective action here is an one-particle irreducible action (1PI) with
all physical excitations already incorporated,
rather than a Wilsonian effective action with a cut-off scale.

This statement has a caveat in the case of Sakai-Sugimoto type models
where the mesons are introduced as degrees of freedom on a probe
brane. What the latter means is that the loop effects of the quark-like
particle are not taken into account by this holographic prescription.
In other words, such a holographic model will at best match with
quenched version of QCD. This is to be expected when $N_c$ is large
and $N_F$ is finite, since the fermion loops would be suppressed by
$N_F/N_c$. It is only when one tries to extrapolate to the real QCD
regime of $N_c=3$ that we must worry about how quenching of the
fermion should be counteracted. However, in this paper, we will
work within the spirit of large $N_c$ QCD, and ignore this issue.

As was studied in depth recently \cite{Hong:2007kx,Hata:2007mb},
the baryon appears in an
entirely different manner. One may recall that, in the
conventional Chiral lagrangian approach, the baryon appears
as a nonperturbative soliton called Skyrmion. In this holographic
and five-dimensional setup, Skyrmion is replaced by
another type of soliton which carries unit Pontryagin
number in the bulk. We will call it an instanton soliton. Furthermore,
the instanton soliton has been shown to shrink to a size
$\sim 1/(M_{KK}\sqrt{\lambda})$ and be localized
at the center of the fifth direction.
An important advantage in the  small soliton size is that
one naturally can resort to an effective field theory language
in the precisely the same sense as the above effective action
of mesons.

For large solitons, which are semi-classical objects, introduction
of an effective field could be a tricky business since we would be trying
to represent a large fluffy objects in terms of point-like quanta
of an effective field. On the other hand, we know from study of dualities
that sometimes one can formulate a theory with soliton in terms of a new
field whose elementary excitation is identified with the soliton.
When would it be justified? It is justified precisely when the
parameters of the original theory approaches a strong coupling
regime so that the size of the soliton becomes smaller that the
typical length scales of the theory.

In this example, we have a soliton  whose Compton size scales as
$1/(\lambda N_c M_{KK})$ and whose soliton size scales as
$1/(\sqrt{\lambda}M_{KK})$. In contrast, the mass scales
of the mesons are fixed at $M_{KK}$. Thus, both the Compton size and
the soliton size of the baryon is much smaller than any of the meson
scale. This tells us is that it is perfectly sensible to
introduce an effective field in place of the  soliton
for the purpose of studying interactions with the meson sector.
On another side of the matter, the classical soliton picture remains robust
since its Compton size is smaller than the soliton size, which allows
us to exploit properties of the classical soliton solution
(whose classical field is made out of mesons) in reading out
interactions of the mesons with the soliton.

In Refs.~\cite{Hong:2007kx,Hong:2007ay,Hong:2007dq}, this program was carried out for the
lowest lying excitation of the soliton, to be identified
with the nucleons with are of isospin 1/2. However, there
is really no reason to truncate to nucleons since the next
excitation, say isospin 3/2 $\Delta$ particles, are not
too heavy compared to the nucleons. The purpose of this note
is to extend this program and read out baryons of
arbitrary isospin and their interactions with mesons
and other baryons.

\section{Baryons as 5D Solitons}

A wrapped D4 brane along the compact $S^4$  corresponds to a
baryon vertex on the five-dimensional spacetime \cite{sakai-sugimoto}, which follows from an argument
originally given by Witten~\cite{witten-baryon}. To distinguish
such D4 from QCD D4's, let us call them  D4'. On their worldvolume
brane we have a Chern-Simons coupling of the form,
\begin{equation}
\mu_4\int C_3\wedge 2\pi\alpha'{d\tilde {\cal A}}=
2\pi\alpha'\mu_4\int dC_3\wedge \tilde {\cal A}
\end{equation}
with D4' gauge field $\tilde { \cal A}$. Since D4' wraps the $S^4$ which
has a quantized $N_c$ flux of $dC_3$, one finds that this term
induces $N_c$ unit of electric charge on the wrapped D4'. The
Gauss constraint for $\tilde {\cal A}$ demands that the net charge
should be zero, however, and the D4' can exist only if $N_c$
fundamental strings end on it. In turn, the other end of
fundamental strings must go somewhere, and the only place it can
go is D8 branes. Thus a D4' wrapping $S^4$ looks like an object
with electric charge with respect to the gauge field on D8. With
respect to the overall $U(1)$ of the latter, which counts the
baryon number, the  charge is $N_c$. Thus,  we may
identify the baryon as wrapped D4' with $N_c$ fundamental strings
sticking onto it.

\subsection{Baryons as ``Small" Instantons with Coulombic Hair}

This wrapped D4'  dissolves into D8 branes and
become an instanton soliton on the latter.\footnote{Usual low energy QCD
picture of baryons as the Skyrmion is directly related to this
instanton picture. As was pointed out by Atiyah and Manton \cite{Atiyah:1989dq}, an
open Wilson line in the presence of an instanton carries the Skyrmion
winding number. Here, the Wilson line along the holographic
direction is nothing but the pion field $U$, completing this
correspondence between the instanton picture and the Skyrmion
picture. From this viewpoint, the instanton soliton can be
thought of as the Skyrmion which is corrected by
the infinite tower of vector and axial vector mesons.
Corrections after including the lightest vector meson only
has been previously considered in Ref.~\cite{vector-skyrmion,Nawa:2007gh}.
However, the full holographic picture seems to change the large $N_c$
nature soliton more profoundly.} The reason for why D4'
cannot dissociate away from D8 is obvious. The D4' has $N_c$
fundamental strings attached, whose other ends are tied to D8.
Moving away from D8 by distance $L$ means acquiring extra mass of
order $N_cL/l_s^2$ due to the increased length of the strings, so
the D4' would have to stay on top of D8 for a simple energetics reason
\cite{Hong:2007kx,Hata:2007mb}.

Once on top of D8's, the D4' is replaced by an instanton
configuration with
\begin{equation}
\frac{1}{8\pi^2}\int_{R^3\times I} \tr F\wedge F=1 \: ,
\end{equation}
where $F$ is the $SU(N_F)$ part of the D8 gauge field strength ${\cal F}$.
This is a well-known consequence of  the Chern-Simons term
on D8,
\begin{equation}
\mu_8\int_{R^{3+1}\times I \times S^4}\, C_{5} \wedge 2\pi^2(\alpha')^2
\tr F\wedge F =\mu_4 \int_{R^{0+1}\times S^4}
C_5\wedge\frac{1}{8\pi^2}\int_{R^3\times I}\,\tr F\wedge F \:,
\end{equation}
which shows that a unit instanton couples to $C_5$ minimally, and
carries exactly one unit of D4' charge.

How about the size of the instanton soliton? Consider the kinetic part of D8 brane action,
compactified on $S^4$, in the Yang-Mills approximation,
\begin{equation}
-\frac14\;\int\sqrt{-g_{4+1}} \;\frac{e^{-\Phi}V_{S^4}}{2\pi (2\pi
l_s)^5} \;\tr {\cal F}_{\hat M\hat N}{\cal F}^{\hat M\hat N}=-\;\int dx^4 dw
\;\frac{1}{4e^2(w)} \;\tr {\cal F}_{MN}{\cal F}^{MN}\:,
\end{equation}
where the unhatted indices are those associated with the flat metric $dx_\mu dx^\mu+dw^2$,
and the electric coupling is $w$-dependent,
\begin{equation}
\frac{1}{e^2(w)}\equiv \frac{8\pi^2 R^3U(w)}{3 (2\pi l_s)^5 (2\pi
g_s)}=\frac{(g_{YM}^2N_c)N_c}{108\pi^3}M_{KK}\frac{U(w)}{U_{KK}}\:.
\end{equation}
Suppose that we have a point-like instanton localized at $w=0$. Its energy
from the Yang-Mills kinetic term would be the standard instanton action,
\begin{equation}
m_B^{(0)}\equiv \frac{4\pi^2}{e^2(0)}=\frac{(g_{YM}^2N_c)N_c}{27\pi}M_{KK}\:.
\end{equation}
For a slightly larger instanton, on the other hand,
$w$-dependence of $e(w)^2$ will induce more energy since
the kinetic term is proportional to $1/e(w)^2$.
For small size parameter $\rho$ such that $\rho M_{KK}\ll 1$, this
extra energy
is\footnote{ The estimate of energy
here takes into account the spread of the instanton density
$D(x^i,w)\sim \rho^4/(r^2+w^2+\rho^2)^4$, but ignores
the deviation from the flat geometry along the four
spatial directions.}
\begin{equation}\label{mass}
\delta m_B^{Pontryagin} \simeq \frac16\, m_B^{(0)}M_{KK}^2\rho^2\:,
\end{equation}
Thus, in the absence of any other effect, the instanton would shrink to
$\rho=0$.

On the other hand, the instanton soliton is really a
representation of a wrapped D4' with $N_c$ fundamental strings
attached. The effect of these fundamental strings are encoded in
the world-volume gauge theory as a Chern-Simons term,
\begin{equation}
\frac{N_c}{24\pi^2}\int \omega_5({\cal A})\:.
\end{equation}
which implies that, for $N_F=2$, the $U(1)$ part of ${\cal A}$
will see a charge density proportional to  the Pontryagin density of the
instanton. Since the Coulomb repulsion favors less and less dense
 charge distribution,
this effect goes to expand the instanton size. More precisely,
the five dimensional Coulomb energy goes as
\begin{equation}\label{C}
\delta m_B^{Coulomb}\simeq \frac12\times \frac{e(0)^2N_c^2}{10\pi^2\rho^2}\:,
\end{equation}
again provided that $\rho M_{KK} \ll 1$.

The competition of the two effects sets the size  to minimize
$\delta m_B^{Coulomb}+\delta m_B^{Pontryagin}$, which is achieved
at \cite{Hong:2007kx,Hata:2007mb}
\begin{equation}\label{size}
\rho_{baryon}\simeq \frac{({2\cdot 3^7\cdot\pi^2/5})^{1/4}}{M_{KK}\sqrt\lambda }\:,
\end{equation}
with the classical mass
\begin{equation}
m_B^{classical}= m_B^{(0)}\times\left(1+\frac{\sqrt{2\cdot 3^5\cdot \pi^2/5}}{\lambda} +\cdots\right)
\end{equation}
For an arbitrarily large 't Hooft coupling limit, the size $\rho_{baryon}$
is significantly smaller than the Compton sizes of the
mesons $\sim 1/M_{KK}$ but much larger than its own Compton size
$1/m_B^{classical}\simeq 27\pi/(M_{KK}\lambda N_c)$ .

Before proceeding further, we should point out that the size of the soliton scales the same
way as $l_s^{warped}$. This tells us that the Yang-Mills Chern-Simons
action we used so far may not be completely reliable. Plugging in the
numbers, we see that the size of the soliton is about four times
larger than $l_s^{warped}$, making it not too small but not large enough
to avoid stringy corrections
either. Consideration of higher order stringy effects will likely
shift the size estimate we have here, making a quantitative correction.
Whether or not we should include these corrections depends on what
we wish to do. It is true that the stringy theory model at hand clearly
demands any such corrections be included. On the other hand, the stringy
holographic QCD model should be reliably dual to ordinary QCD only in the
energy scale far below $M_{KK}$ anyway, yet have successfully reproduced
certain behaviors of QCD around $M_{KK}$ as well. How and why of this are
hardly clear for this model, nor is it clear for any other holographic
QCD. In this sense, the guiding principle is lost once we begin to
consider any massive objects, as far as we are interested in emulating
real QCD.

With this uncertainty in mind, we will try to proceed without worrying
about such stringy corrections.  A good news is, though, that, in
what follows from here, where we effectively consider soft scattering
processes involving meson, this problems is much less acute. Even though
the mass scale and the length scale of the hadrons are dangerously
high and small, actual physical process to be considered are such
that the  momentum transfer is typically no larger than $M_{KK}$ and
more like $f_\pi$. When we compute corrections to $\rho_{baryon}$ by
whatever higher order effect, all we have to do  is to replace
our size parameter by the corrected one in what follows, and the rest
is intact.

\subsection{Quantization}

If the soliton size is small, physics near a soliton
located at $w=0$ retains the approximate symmetry of $R^{4+1}$. The
deviation from this symmetry is an important ingredient
that enters the size estimate of the soliton and also
must be considered carefully for reading out interactions
between baryons and mesons. However, we will temporarily
ignore this deviation since here we concentrate on the
counting of the quantum states, for which the approximate
$R^{4+1}$ Minkowskian invariance can be very useful, and the
result robust under the deviations. Matching of quantized soliton
with the baryon is easiest when the number of flavor is two.
{}From this point on, we specialize to the case of
$N_F=2$.

In order to set up an effective action of baryons, it is
important to understand what kind of quantum states emerges from
quantizing these solitons. Usual $SU(2)$ instanton in flat $R^4$
carries eight collective coordinates, four translational ones, three global
$SU(2)$ rotations, and one overall size. Of these, the last is not
a moduli direction for our instanton, but the other seven are all
from broken symmetry and thus remain flat. To elevate
the instanton soliton to a point-like object, i.e. a quantum of
an effective field, we must quantize  some of these collective coordinates
and produces representations under the symmetry of the moduli space.

The approximate Lorentz group at hand is $SO(4,1)$, to be broken to
$SO(3,1)$ by the curvature effect etc. The approximate
 little group for massive particle is $SO(4)_{R^4}=SU(2)_+\times SU(2)_-$.
Classical self-dual instanton rotates nontrivially under one of
the two factors, say $SU(2)_+$, while classical anti-instanton
rotates under $SU(2)_-$. Instantons also gets rotated by the
global gauge rotation $SU(N_F=2)$,
\begin{equation}
F\quad\rightarrow\quad S^\dagger FS \:,
\end{equation}
with special unitary matrices $S$. The collection of $S$ spans the
$SU(2)$ manifold, or equivalently ${\mathbf S}^3$, but since $S$ and $-S$
rotates the solution the same way the moduli space is naively
${\mathbf S}^3/Z_2$. However at quantum level, we must consider states
odd under this $Z_2$ as well, so the moduli space is ${\mathbf S}^3$.
Then, its quantization is a matter of finding eigenstates
of free and nonrelativistic nonlinear sigma-model onto ${\mathbf S}^3$
\cite{ANW,Witten:1983tx,Finkelstein:1968hy}.

$S$ itself admits $SO(4)$ symmetry action of its own, which can be
written as
\begin{equation}
S\quad\rightarrow \quad U S V^\dagger \;.
\end{equation}
Because of the way the spatial indices are locked with the
gauge indices, these two rotations are each identified as
the gauge rotation, $SU(2)_I$, and half of spatial rotation,
$ SU(2)_+$. For each factor, we have a triplet of symmetry
operators, $I_{1,2,3}$ and $J^{(+)}_{1,2,3}$, respectively.

Eigenstates on ${\mathbf S}^3$ are nothing but the
angular momentum eigenstates under $I$'s and $J^{(+)}$'s,
conventionally denoted as
\begin{equation}
\vert s:p,q\rangle \;,
\end{equation}
with the eigenvalues $I^2=s(s+1)=(J^{(+)})^2$, $I_3=p$, and $J_3^{(+)}=q$. As is well-known,
$I^2$ and $J^2$ eigenvalues are always equal on ${\mathbf S}^3$, so that spin $s$ baryons
are always in isospin $s$ representation as well. For even $s$,
$D^{(s)}$'s are even under the $Z_2$, and for odd $s$, $D^{(s)}$'s are
odd under $Z_2$.

The simplest way to represent these eigenstates
as wavefunctions on ${\mathbf S}^3$
are to use the Cartesian representation of the Euler angles  as
\begin{equation}
S=S(\xi)=\xi_4+i\xi_a\tau_a\:,\quad \xi^2=1 \;,
\end{equation}
for the $2\times 2$ Pauli matrices $\tau_i$'s. The eigenstates
have a well-known representation in the coordinate basis as
functions on ${\mathbf S}^3$,
\begin{equation}
D^{(s)}_{pq}(\xi)=\langle \xi \vert s:p,q\rangle \;,
\end{equation}
and we can further choose the basis for $\xi$'s such that
\begin{equation}
D^{(s)}_{ss}(\xi)=\sqrt{\frac{2s+1}{2}}\:\frac{1}{\pi}\:(\xi_1+i\xi_2)^{2s} \;.
\end{equation}
The spin and isospin operators
are realized as differential operators
\begin{eqnarray}
I_a&\rightarrow &{\cal I}_a\equiv-\frac{i}{2}
\left(\epsilon_{abc}\xi_b\partial_c - \xi_4\partial_a+\xi_a\partial_4\right) \;,\nonumber\\
J_a^{(+)}&\rightarrow &{\cal J}_a^{(+)}\equiv
-\frac{i}{2}\left(\epsilon_{abc}\xi_b\partial_c + \xi_4 \partial_a-\xi_a\partial_4\right) \;.
\end{eqnarray}
It is easy to show that ${\cal I}^2=({\cal  J}^{(+)})^2 $ holds as the consistency would require.

One can proceed exactly the same manner for anti-instantons, where
$SU(2)_+$ is replaced by $SU(2)_-$. Therefore,
under $SU(2)_I\times SO(4)_{R^4}=SU(2)_I\times SU(2)_+\times SU(2)_-$,
quantized instantons are in
\begin{equation}
(2s+1;2s+1;1)
\end{equation}
while quantized anti-instantons are in
\begin{equation}
(2s+1;1;2s+1)
\end{equation}
Theoretically possible values for $s$ are integers or half-integers. However,
we are mainly interested in fermionic baryons, and will subsequently
consider the case of half-integral $s$'s.\footnote{For $N_F$ larger
than two, the half-integral spin of the baryons should follows immediately
whenever $N_c$ is odd, in a manner  similar to the Skyrmion case \cite{Witten:1983tx}.}

Before closing, let us note that the instanton and
anti-instanton can be naturally thought of as particle/anti-particle
pairs. The representation under the little group reflects this as well.
Later, we will introduce an effective field whose elementary quanta
are these particles and anti-particles. Due to CPT, the particle
and the anti-particle always come together, and we expect to find an
effective field that produce excitations that belong to
\begin{equation}
(2s+1)_{SU(2)_I}\otimes\left((2s+1;1)\oplus (1;2s+1)\right)_{SO(4)_{R^4}}
\end{equation}
on-shell.
This sets the table for extracting quantum baryons out of
the classical instanton soliton.

\section{Effective Field Theory of the Instanton Soliton}

In this section, we wish to introduce an effective field
whose elementary quanta are the quantized
baryons of the previous section. We  introduce the
field content for any given isospin and propose an
effective action of such baryon fields interacting
with the gauge field, a.k.a., mesons. In next section,
we will derive the proposed effective action by
generalizing a method originally due to Adkins, Nappi, and Witten\cite{ANW}
and also adapted for holographic Nucleons in Refs.~\cite{Hong:2007kx,Hong:2007ay}.

\subsection{Higher Spin Fields in Five Dimensions}

We learned that quantization of instanton
soliton leads to quantum states with isospin and spin related.
For isospin $s$, the quantized instantons are in $(2s+1,1)$ and
the quantized anti-instantons are in $(1,2s+1)$ under the
little group $SO(4)=SU(2)_+\times SU(2)_-$. For $s=1/2$, they
combine into a Dirac field with a single spinor index, so one
might think that, for $s=3/2$, the relevant field is the
Rarita-Schwinger field. However, this would be true only if we
are working in four dimensions. The Rarita-Schwinger field in
five dimensions produces six particle and six anti-particle
degrees of freedom. More precisely they are in the representations
$(3,2)+(2,3)$ under the little group.

The right choice  is the higher
spin field with multiple spin indices, completely
symmetrized,
\begin{equation}
\Psi_{A_1A_2\cdots A_{2s}}=\Psi_{(A_1A_2\cdots A_{2s})} \;,
\end{equation}
where the spin index $A$ runs from 1 to 4. We consider
half-integral $s$, since real QCD admits only those.
Consider a free equation of motion
\begin{equation}
\gamma^M_{A_1B}\partial_M \Psi_{BA_2\cdots A_{2s}}=m \Psi_{A_1A_2\cdots A_{2s}}\;,
\end{equation}
where  the Dirac matrices act on the first spin index only.
The Dirac operator squares to $\nabla^2$, so we find the equation of motion
implies the usual on-shell condition $p^2+m^2=0$, which of course gives
$E^2=m^2$ in the rest frame. This further imposes the condition on the plane-wave spinors
in the rest frame as
\begin{equation}
\mp i\gamma^0\Psi= \Psi
\end{equation}
so that particles and anti-particles
correspond to $-i\gamma^0$ eigenstates with eigenvalues $\pm 1$.
Due to the symmetrized spin indices, this implies that
a $-i\gamma^0$ eigenstate must have the same ``chirality" for
all $2s$ indices.

On the other hand, $\Gamma\equiv \gamma^1\gamma^2\gamma^3\gamma^4=-i\gamma^0$,
so particles and anti-particles are, respectively,
chiral and anti-chiral under the little group $SO(4)_{R^4}=SU(2)_+\times SU(2)_-$.
With the
following choice for Dirac matrices for $m,n=0,1,2,3,4$,
\begin{equation}
\gamma^0=\left(\begin{array}{rr} i & 0 \\ 0 & -i\end{array}\right),\quad
\gamma^i=\left(\begin{array}{cc} 0 & \sigma_i \\ \sigma_i & 0\end{array}\right),\quad
\gamma^4=\left(\begin{array}{rr} 0 & i \\ -i & 0\end{array}\right) , \label{gamma}
\end{equation}
 particles are encoded in
\begin{equation}
\Psi_{(A_1A_2\cdots A_{2s})}\:,\quad A_i=1,2 \;,
\end{equation}
whose individual spinor indices $A_i=1,2$ belong to doublets under $SU(2)_+$.
These particles are clearly in the spin $s$ representation of $SU(2)_+$.
Likewise, anti-particles
\begin{equation}
\Psi_{(A_1A_2\cdots A_{2s})}\:,\quad A_i=3,4 \;,
\end{equation}
are in the spin $s$ representation under $SU(2)_-$.

The upshot is that the propagating degrees of freedom form
\begin{equation}
\left((2s+1,1)\oplus(1,2s+1)\right) \;,
\end{equation}
under the little group ${SO(4)_{R^4}}$.  After
elevated to the isospin $s$ under $SU(2)_I$, this spinor fields is
then capable of
reproducing particle contents of quantized instantons and anti-instantons.
Since the spin and the isospin are locked, the field representing the
quantized instanton and anti-instanton also carry the flavor  $SU(2)$
indices
\begin{equation}
\Psi^{\epsilon_1\cdots \epsilon_{2s}},\quad \epsilon_i=1,2 \;.
\end{equation}
For the case of $s=1/2$, the authors of \cite{Hong:2007kx,Hong:2007ay}
wrote down a relativistic
field theory involving the nucleons and the gauge field.

For $s\ge 3/2$, however, this is easier said than done.  For four
dimensions, Rarita-Schwinger field does the trick for $s=3/2$ but
we cannot use this  in five dimensions due to a different spin
content. The only
sensible way out, at least until we know better, is to employ the
nonrelativistic approximation. This is well justified in the large $\lambda N_c$
limit, since the mass of the instanton scales as $\lambda N$. Thus, instead
of working with fully relativistic four-component spinor notations,
we will split it to particle and anti-particles as
\begin{eqnarray}
-i\gamma^0\Psi^{\epsilon_1\cdots \epsilon_{2s}}= \Psi^{\epsilon_1\cdots \epsilon_{2s}}
&\rightarrow & {\cal U}_{\alpha_1\cdots \alpha_{2s}}^{\;\epsilon_1\cdots \epsilon_{2s}},\quad \alpha_i=1,2
\;, \nonumber \\
i\gamma^0\Psi^{\epsilon_1\cdots \epsilon_{2s}}= \Psi^{\epsilon_1\cdots \epsilon_{2s}}
&\rightarrow & {\cal V}_{\dot\beta_1\cdots \dot\beta_{2s}}^{\;\epsilon_1\cdots \epsilon_{2s}},\quad \dot\beta_i=1,2
\;.
\end{eqnarray}

\subsection{Nonrelativistic Lagrangian}

In order to see the interaction terms giving rise to instanton tails
it is sufficient to look for the nonrelativistic limit of higher spin
theories. We just have to look for the interactions of the particle
rather than antiparticle. As above, we denote
\begin{equation}
{\cal U}_{\alpha_1\cdots \alpha_{2s}}^{\;\epsilon_1\cdots \epsilon_{2s}}
\end{equation}
be the positive energy components or particle components where
$a^i$'s, $\alpha^i$'s take values of 1,2. Indices $a^i$'s, $\alpha^i$'s
are all symmetrized.
The minimal Lagrangian of the nonrelativistic limit of the particle of
spin $s$ is given by the usual Schrodinger type
\begin{equation}
S_0=\int dt d^4x\;\left[\sum_s \left( i{\cal U}_s^{\dagger}\frac{\partial}{\partial t}{\cal U}_s
+\sum_{m=1}^4 \frac{1}{2m(s)}{\cal U}_s^{\dagger}(\partial_m-i{\cal A}_m)^2 {\cal U}_s\right)\right]
\;,
\end{equation}
where ${\cal U}_s$ denotes a field with spin $s$.
Mass of the isospin $s$ baryon is denoted as $m(s)$.
Please see Hata et.al. \cite{Hata:2007mb} for an explicit
formula of excited baryon mass.
In the kinetic term, the $U(2)$ gauge
field enters in the following combination
\begin{equation}
{\cal A}_m=N_cA_{U(1)}
+A^{(s)}\;,
\end{equation}
where $A^{(s)}$ is the isospin $s$ representation of the
non-Abelian $SU(2)$ part of the gauge field.

On the other hand, we anticipate additional couplings to the $SU(2)$
field strength in much the same way for $s=1/2$.
The logic goes as follows. The above minimal interaction tells the
gauge field to generate long-range Coulomb field in response to
the electric charges on the soliton. However, the instanton and
anti-instanton are
characterized by the self-dual and anti-self-dual magnetic fields
whose power-like tail is determined by
$\rho_{baryon}^2$. Note that this magnetic field goes as
of $1/r^4$,  one more power of $1/r$ than the Coulomb field.
When we replace the quantized instanton by a field,
we must somehow incorporate this aspect of the soliton to re-emerge from the
equation of motion, just as the Coulomb field emerges naturally
from the  minimal coupling.  For the case of $s=1/2$, it was shown
in \cite{Hong:2007kx, Hong:2007ay}  that a direct coupling to the field strength $F=dA+iA^2$ to
a bilinear of the spinor emulates this long range behavior of the
quantized soliton. We wish to generalize this to arbitrary $s$.

The proposal for these additional interaction terms are roughly
\begin{eqnarray}
& &  ({\cal U}^{\epsilon'\epsilon_2 \cdots
  \epsilon_{2s}}_{\beta \alpha_2 \cdots \alpha_{2s}})^*
  (\gamma^0\gamma^{KN})^{\beta \beta'} F^{\epsilon'\epsilon}_{KN}\,\,
{\cal U}^{\epsilon\epsilon_2 \cdots
  \epsilon_{2s}}_{\beta' \alpha_2 \cdots \alpha_{2s}} \;,
  \end{eqnarray}
between baryons of the same isospin, and
  \begin{eqnarray}
&& ({\cal U}^{\epsilon_1\epsilon_2 \cdots
  \epsilon_{2s}}_{\alpha_1\alpha_2 \cdots \alpha_{2s}})^*
  (\gamma^0 C\gamma^{KN})^{\beta \beta'} (\tau_{2}F_{KN})^{\epsilon \epsilon'}\,\,
{\cal U}^{\epsilon\epsilon'\epsilon_1 \cdots
  \epsilon_{2s}}_{\beta \beta'\alpha_1 \cdots \alpha_{2s}} \;,
\end{eqnarray}
between baryons of different isospins. Here, the spinor indices runs
over $1,2$ only (since the nonrelativistic spinors are of two-components),
even though we kept the notation of 5-d gamma matrices on purpose to
indicate  possible relativistic origins of such interactions.
The charge conjugation matrix ${\cal C}$ satisfies
\begin{equation}
(\gamma^{MN})^T=-{\cal C}\gamma^{MN}{\cal C}^{-1} \;,
\end{equation}
and is in our convention
\begin{equation}
{\cal C}=\left(\begin{array}{cc} \sigma_2 & 0 \\ 0 & -\sigma_2\end{array}\right) \;.
\end{equation}
Finally, the electromagnetic field $F^{\epsilon\epsilon'}_{kn}$ is
\begin{equation}
F^{\epsilon\epsilon'}_{KN}\equiv \Sigma_{a=1}^3 F^a_{KN}\frac{\tau_a^{\epsilon\epsilon'}}{2} \;,
\end{equation}
and, for these $SU(N_F=2)$  gauge indices, $\tau_2$ plays the same role as
${\cal C}$ does for the spinor indices.

Even though we are writing down a nonrelativistic action, it is
important to keep in mind that there should be a fully Lorentz
invariant dynamics. Once we show that the particle interaction gives
rise to instanton configurations, the antiparticle
interaction should follow automatically. With this in mind,
let us write these terms in the honest  two-component notations appropriate
for the nonrelativistic spinors.
With the convention of the gamma matrices (\ref{gamma}),  the
interaction terms involving the magnetic fields, $F_{ij}$ and $F_{4i}$, are
\begin{eqnarray}
S_I^{magnetic}&=&-\sum_s \frac12 h_s F_{ij}^a\epsilon_{ijk} ({\cal U}^{\epsilon'\epsilon_2 \cdots
  \epsilon_{2s}}_{\beta' \alpha_2 \cdots \alpha_{2s}})^*
  \sigma_k^{\beta' \beta} \tau^{\epsilon'\epsilon}_a
{\cal U}^{\epsilon\epsilon_2 \cdots
  \epsilon_{2s}}_{\beta \alpha_2 \cdots \alpha_{2s}}  \nonumber \\
& &-\sum_s h_s F_{4i}^a  \,\, ({\cal U}^{\epsilon'\epsilon_2 \cdots
  \epsilon_{2s}}_{\beta' \alpha_2 \cdots \alpha_{2s}})^*
  \sigma_i^{\beta' \beta} \tau^{\epsilon'\epsilon}_a
{\cal U}^{\epsilon\epsilon_2 \cdots
  \epsilon_{2s}}_{\beta \alpha_2 \cdots \alpha_{2s}}  \nonumber \\
& & -\sum_s\frac12 k_s F_{ij}^a \epsilon_{ijk} ({\cal U}^{\epsilon_1\epsilon_2 \cdots
  \epsilon_{2s}}_{\alpha_1\alpha_2 \cdots \alpha_{2s}})^*
  (\sigma_2\sigma_k)^{\beta \beta'} (\tau_2\tau_a)^{\epsilon\epsilon'}
{\cal U}^{\epsilon\epsilon'\epsilon_1 \cdots
  \epsilon_{2s}}_{\beta\beta' \alpha_1 \cdots \alpha_{2s}}
  \nonumber \\
& & -\sum_s k_s F_{4i}^a  ({\cal U}^{\epsilon_1\epsilon_2 \cdots
  \epsilon_{2s}}_{\alpha_1\alpha_2 \cdots \alpha_{2s}})^*
  (\sigma_2\sigma_i)^{\beta \beta'} (\tau_2\tau_a)^{\epsilon\epsilon'}
{\cal U}^{\epsilon\epsilon'\epsilon_1 \cdots
  \epsilon_{2s}}_{\beta\beta' \alpha_1 \cdots \alpha_{2s}} \;, \label{magneticterm}
\end{eqnarray}
where $i,j,k=1 \cdots 3$.
With the usual t'Hooft symbol, this can be written as
\begin{eqnarray}
S_I^{magnetic}&=& -\frac12\sum_s h_s F_{mn}^a\bar{\eta}^b_{mn}
({\cal U}^{\epsilon'\epsilon_2 \cdots
  \epsilon_{2s}}_{\beta'\alpha_2 \cdots \alpha_{2s}})^*
  \sigma_b^{\beta' \beta} \tau_a^{\epsilon'\epsilon}
{\cal U}^{\epsilon\epsilon_2 \cdots
  \epsilon_{2s}}_{\beta \alpha_2 \cdots \alpha_{2s}} \nonumber \\
& & -\frac12\sum_s k_s  F_{mn}^a\bar{\eta}^b_{mn} ({\cal U}^{ \epsilon_1\epsilon_2 \cdots
  \epsilon_{2s}}_{\alpha_1\alpha_2 \cdots \alpha_{2s}})^*
  (\sigma_2\sigma_b)^{\beta \beta'} (\tau_2\tau_a)^{\epsilon\epsilon'}
{\cal U}^{\epsilon\epsilon'\epsilon_1 \cdots
  \epsilon_{2s}}_{\beta\beta' \alpha_1 \cdots \alpha_{2s}} \;.\label{interaction}
\end{eqnarray}
The presence of the anti-self-dual 't Hooft symbol $\bar \eta^a_{mn}$
indicates that above interaction terms will source the smeared-out instanton
field.

There will be an electric analog of these terms, $S_I^{electric}$,
involving the electric field strengths $F_{0m}$ instead of the magnetic
field strength $F_{mn}$. These electric couplings cannot be derived
from the soliton structure, but must be rather inferred via Lorentz
invariance from the magnetic ones. Here, we chose not to display them
explicitly.

%Further identification of this source as the quantized
%instanton requires an understanding of the expectation values involving
%$\sigma_k\tau_m$. For the first class,
%this is precisely the case and the bilinear evaluates the expectation
%value of these as operators for each spin/isospin $s$ field.
%For the second class, the expectation value is between states of
%different total spin, so $\sigma_k\tau_m$ is used to reduce total
%spin by one, instead. In the next part, we will show these terms are
%precisely what are needed to make the effective theory consistent with
%the instanton picture of the baryons.

\subsection{Relativistic Origins}

As an aside, let us note that, as far as the interaction terms go,
we have an obvious relativistic completion.
When we proposed the nonrelativistic effective lagrangian,
we implicitly assumed this underlying relativistic structure.
In particular, CPT invariance is enforced. Even though the baryons
are extremely heavy in the large $\lambda N_c$ limit, their dynamics
must respect the Lorentz invariance. The difficulty involved in
formulating a fully relativistic action is with the kinetic
terms and constraints, rather than with interactions.

In this spirit, we note that term in $S_I^{magnetic}$ and $S_I^{electric}$
would follow from the following structures,
\begin{eqnarray}
h_s'\bar\Psi^{(s)}(F\Psi^{(s)})
+f_{s}\bar\Psi^{(s)}(F\cdot\Psi^{(s+1)}) \;,
\end{eqnarray}
in terms of the relativistic spinor $\Psi$'s.
The  contraction in the second terms is defined as
\begin{equation}
(F\cdot\Psi^{(s+1)})_{A_1A_2\cdots A_{2s}}^{\epsilon_1\cdots \epsilon_{2s}}\equiv
 (\tau_2F_{MN})^{\epsilon\epsilon'}({\cal C}\gamma^{MN})_{BB'}
\Psi_{BB'A_1A_2\cdots A_{2s}}^{\epsilon\epsilon'\epsilon_1\cdots \epsilon_{2s}} \;,
\end{equation}
which lowers the isospin and the spin representations.
It is not difficult to convince oneself that these two are the only
possible fermion bilinears with direct couplings to the field strength.
The resulting Lorentz-covariant form of the Yang-Mills equations is
\begin{eqnarray}
D^M{F}_{MN}^{\epsilon\epsilon'}=\cdots
&+&\sum_s h_s'
D^K\left(\gamma_{KN}^{BB'}
\bar \Psi^{\epsilon'\epsilon_2\cdots\epsilon_{2s}}_{BA_2\cdots A_{2s}}
\Psi^{\epsilon \epsilon_2\cdots \epsilon_{2s}}_{B'A_2\cdots A_{2s}}\right)
\nonumber\\
&+&\sum_s k_s D^K\left(({\cal C}\gamma_{KN})^{BB'}\tau_2^{\nu\epsilon}
\bar \Psi^{\epsilon_1\cdots\epsilon_{2s}}_{A_1\cdots A_{2s}}
\Psi^{\nu\epsilon' \epsilon_1\cdots \epsilon_{2s}}_{BB'A_1\cdots
A_{2s}}\right)\;,
\end{eqnarray}
where the ellipsis denote the baryon current that account for the
$N_c$ charge of the baryon.

There are two notable differences between the relativistic and the
nonrelativistic expressions. First is that we displayed in Eq.~(\ref{interaction})
only the couplings to the magnetic field, whereas the relativistic
form includes the same type of couplings to the electric field.
As far as the derivation of the coupling goes, only the magnetic
one can be derived since it comes from the self-dual magnetic
field strength of the instanton. The electric one has to be
there simply because Lorentz symmetry relates the two.

Secondly, we have $h_s'$ in place of $h_s$  because there is
another relativistic source of the same magnetic terms. These
magnetic terms are five-dimensional analog of non-anomalous magnetic moment term
of four-dimensional Dirac field, and arise when we expand the minimal coupling
terms in terms of nonrelativistic spinors.
In \cite{Hong:2007kx,Hong:2007ay}, this  correction was ignored since in the large $N_c$
limit, it represents a second order correction of order $\sim 1/m_B$. However, in the
extrapolation to finite $N_c$, this correction, if kept, could be
comparable or even larger than the leading term. This makes the
extrapolation procedure somewhat ambiguous. In this note, we will
stick to large $N_c$ limit, and ignore this problem.

\section{Derivation of The  Interaction Terms}

In the previous section, we speculated on possible interactions
between baryons and mesons. While we discussed direct couplings
to the field strengths in five dimensions, we are yet to show that the structure
we gave is indeed the right one. In this section, we will generalized
the work in \cite{Hong:2007kx,Hong:2007ay}, and show that these
couplings are inevitable consequences
of the instanton origin of the baryon. As a by-product, we also
compute the strength of the couplings at origin, $w=0$.

The strategy for this goes as follows. When the instanton is
quantized, its classical gauge field configuration is replaced by
its expectation values as
\begin{equation}
F\;\rightarrow\;\langle\langle S^\dagger FS\rangle\rangle \;,
\end{equation}
where $\langle\langle\cdots\rangle\rangle$ means taking expectation
value on the collective coordinate wavefunctions.
Componentwise, with the explicit $SU(2)$ generators $\tau_a/2$, we have
\begin{equation}\label{smeared}
F^a\;\rightarrow\; \langle\langle \tr \left[\tau_a S^\dagger \frac{\tau_b}{2} S\right]\rangle\rangle F^b
\;,
\end{equation}
so that the quantum smearing out  of the  instanton gauge field
is determined entirely by the expectation value of the quantities,
\begin{equation}
\Sigma_{ab} \equiv \tr\left[\tau_a S^\dagger
\frac{\tau_b}{2}S \right] \;.
\end{equation}
If and only if any unit quanta of the baryon field can emulate such
a smeared-out long range field, the effective baryon field theory
would make sense.

Fortunately, the spin-isospin-locked nature of the instanton
wavefunction allows us to translate this expectation value
in terms of bilinears of the baryon field. The simplest of such
relation was found by Adkins-Nappi-Witten for isospin 1/2 case
in the context of quantized Skyrmions\cite{ANW}, which is
\begin{equation}\label{anw2}
\langle\langle 1/2:p',q'\vert\,\tr \left[\tau_a S^\dagger \frac{\tau_b}{2} S\right]\vert 1/2:p,q\rangle\rangle
=-\frac{4}{3}\:\langle 1/2:p',q'\vert J_a^{(\pm)} I_b\vert
1/2:p,q\rangle\:.
\end{equation}
On the left hand side, we have overlap
integrals of functions on $S^3$,
\begin{equation}
\langle\langle s:p',q'\vert\, \tr \left[\tau_a S^\dagger \frac{\tau_b}{2} S\right]\vert s:p,q\rangle\rangle
\equiv\int_{{\mathbf S}^3}
\left(D^{(s)}_{p'q'}(\xi)\right)^*D^{(s)}_{pq}(\xi) \:\Sigma_{ab}(\xi)\: ,
\end{equation}
whereas the quantity on the right hand side is the usual matrix elements of
angular momentum operators. We can conveniently represent the right hand side as
\begin{equation}
4\:\langle 1/2:p',q'\vert J_a^{(\pm)} I_b\vert
1/2:p,q\rangle\:\equiv ({\cal U}(1/2:p',q')^{\epsilon'}_{\beta'})^*
  \sigma_a^{\beta' \beta} \tau_b^{\epsilon'\epsilon}
{\cal U}(1/2:p,q)^{\epsilon}_{\beta} \;,
\end{equation}
in terms of a two-component spinor field ${\cal U}$ in the
isospin 1/2 representation.

Note that the effective action of ${\cal U}$ in Eq.~(\ref{interaction}) we proposed is such
that ${\cal U}$ bilinear sources the five-dimensional gauge field as
\begin{equation}\label{ymeom}
(\nabla\cdot F)_m^a\sim \nabla_n\left(\bar\eta^b_{nm}{\cal U}^\dagger(\sigma_b\tau^a){\cal U}\right)+\cdots.
\;.
\end{equation}
Thanks to the above identity (\ref{anw2}), this Yang-Mills field equation implies that
a ${\cal U}$ particle state  will have a long range
tail of gauge field that looks exactly like a smeared out instanton
of Eq.~(\ref{smeared}), as long as we match the
precise mapping between the spinor states and the
quantized instanton wavefunctions. This proves that the couplings in the
previous section is indeed exactly the right ones for the
spinor ${\cal U}$ to be interpreted as the baryon effective
field for $s=1$. In the remainder of this section, we will show how
this  generalizes to all isospins.

Note that together with the obligatory minimal coupling,
this fixes the effective interaction of the baryon with
the meson sector uniquely up to dimension six operators in five dimensions.
The coupling strength is then determined by making
sure the Yang-Mills solution to this equation has exactly the
same size as the smeared out instanton. The subsequent
reduction to four dimensions  generates an infinite number of
coupling constants between mesons and baryon current, as we will
see shortly.

\subsection{Identities for Isospin-Preserving Processes}

For $s>1/2$, the identity (\ref{anw2}) can be generalized to
general integer $s$, as
\begin{equation}\label{newanw}
\langle\langle s:p',q'\vert\Sigma_{ab}\vert s:p,q\rangle\rangle
\:
=-C_0(s)\langle s:p',q'\vert J_a I_b\vert s:p,q\rangle \: ,
\end{equation}
for arbitrary $s$ and $-s\le p,q,p',q'\le s$ with
\begin{equation}
C_0(s)=\frac{1}{s(s+1)} \;.
\end{equation}
To show this, let us start with the simplest case of $p=q=p'=q'=s$.
For this, it is relatively easy to show that
\begin{equation}\label{top}
\int_{{\mathbf S}^3}\vert D^{(s)}_{ss}\vert^2 \:\Sigma_{ab}\:
=-C_0(s)\langle s:s,s\vert J_a^{(\pm)} I_b\vert s:s,s\rangle \;,
\end{equation}
holds for all $3\times 3 $ choices of $(k,m)$ and for arbitrary integer $s$.
The right hand side is obvious: all cases except $k=m=3$
vanish, and for $k=m=3$ we find $-s^2/s(s+1)=-s/(s+1)$. An explicit
computation of the integral on the left hand side is also
straightforward and produces the same result.

Further generalization follows from the fact that the operators on the two sides
transform the same way under $SU(2)_I\times SU(2)_\pm$. Recalling
how $S$ transforms, we see that
\begin{eqnarray}
\Sigma_{km}\quad\rightarrow\quad\Sigma'_{ab}
&=&\tr\left[\tau_a\;(V S^\dagger U^\dagger)\;\frac{\tau_b}{2}\;(US V^\dagger) \right]\nonumber\\
&=&\tr\left[(V^\dagger \tau_a V) \;S^\dagger \;(U^\dagger\frac{\tau_b}{2}\,U)\;S \right]\nonumber\\
&=&\hat V_a^{\;\;\;c}\hat U_b^{\;\;\;d}\Sigma_{cd} \;,
\end{eqnarray}
where $\hat V$ and $\hat U$ are the $3\times 3$ matrix
representation of $V$ and $U$. Thus, $\Sigma_{km}$
transform under the ${\cal I}$'s and ${\cal J}$'s
exactly as the operators $J_k^{(\pm)}I_m$ would transform
under $I$'s and $J^{(\pm)}$'s.
Next, consider
\begin{equation}\label{top2}
\langle\langle s:s,s\vert\Sigma_{ab}\vert s:p,q\rangle\rangle
\:
=-C_0(s)\langle s:s,s\vert J_a I_b\vert s:p,q\rangle \:,
\end{equation}
for all $-s\le p,q\le s$.  It is easy to show that, of these, the only
nonvanishing expressions are those with $p,q \ge s-1$.
%say,
%\begin{eqnarray}\label{anw4}
%\langle\langle s:s,s\vert\Sigma_{3+}\vert s:s,s-1\rangle\rangle
%&=&-\frac{1}{s(s+1)}\langle s:s,s\vert J_3 I_+\vert s:s,s-1\rangle \nonumber \\
%\langle\langle s:s,s\vert\Sigma_{+3}\vert s:s-1,s\rangle\rangle
%&=&-\frac{1}{s(s+1)}\langle s:s,s\vert J_+ I_3\vert s:s-1,s\rangle \\
%\langle\langle s:s,s\vert\Sigma_{++}\vert s:s-1,s-1\rangle\rangle
%&=&-\frac{1}{s(s+1)}\langle s:s,s\vert J_+ I_+\vert s:s-1,s-1\rangle \nonumber \: .
%\end{eqnarray}
Taking $p=s,q=s-1$, for instance, we see that the left hand side reduces
\begin{eqnarray}\label{anw4}
&&\langle\langle s:s,s\vert\Sigma_{3+}\vert s:s,s-1\rangle\rangle \nonumber \\
&=&\frac{1}{\sqrt{2s}}\langle\langle s:s,s\vert\Sigma_{3+}{\cal I}_- \vert s:s,s\rangle\rangle\nonumber \\
&=&\frac{1}{\sqrt{2s}}\langle\langle s:s,s\vert 2\Sigma_{33} \vert s:s,s\rangle\rangle \;,
\end{eqnarray}
which is exactly mirrored by the left hand side because $[I_+,I_-]=2I_3$
and $\langle s:s,s\vert I_- =0$. So, the first identity in (\ref{top2}) follows
from (\ref{top}). The remaining two can be shown likewise. Continuing in this
fashion, the rest of the identity in Eq.~(\ref{newanw}) follows automatically.

 The right hand side is more conveniently
represented in terms of the nonrelativistic spinor of
the previous section as
\begin{eqnarray}\label{newanw2}
&&\langle\langle s:p',q'\vert\Sigma_{ab}\vert s:p,q\rangle\rangle \: \nonumber\\
&=&-C_0(s)\,{{\cal U}(s:p',q')^\dagger} (J_a\otimes I_b {\cal
U}(s:p,q)) \nonumber \\
&=&-C_0(s)\times s^2\times({\cal U}(s:p',q')^{\epsilon'\epsilon_2 \cdots
  \epsilon_{2s}}_{\beta'\alpha_2 \cdots \alpha_{2s}})^*
  \sigma_a^{\beta' \beta} \tau_b^{\epsilon'\epsilon}
{\cal U}(s:p,q)^{\epsilon\epsilon_2 \cdots
  \epsilon_{2s}}_{\beta \alpha_2 \cdots \alpha_{2s}}\nonumber \\
  &=&-\frac{s}{s+1}\times({\cal U}(s:p',q')^{\epsilon'\epsilon_2 \cdots
  \epsilon_{2s}}_{\beta'\alpha_2 \cdots \alpha_{2s}})^*
  \sigma_a^{\beta' \beta} \tau_b^{\epsilon'\epsilon}
{\cal U}(s:p,q)^{\epsilon\epsilon_2 \cdots
  \epsilon_{2s}}_{\beta \alpha_2 \cdots \alpha_{2s}} \;,
\end{eqnarray}
where the operators $J$ and $I$ acting on ${\cal U}$ are of
understood to be in the spin (isospin) $s$ representation. Note that
the last expression, up to an overall numerical factor,
is precisely the first of  the two fermion bilinears
that appeared in Eq.~({\ref{interaction})

\subsection{Generalizing to Isospin-Changing Processes}

Note that $\Sigma_{km}$ are themselves spin one
wavefunctions on $S^3$. With the non-Hermitian choice of
basis $\tau_\pm=(\tau_1\pm i \tau_2)/\sqrt{2}$, we find
\begin{eqnarray}
\Sigma_{++}(\xi)&=&\sqrt{{2\pi^2}/{3}}\; D^{(1)}_{11}(\xi)=(\xi_1+i\xi_2)^2\;,\nonumber\\
\Sigma_{3+}(\xi)&=&\sqrt{{2\pi^2}/{3}}\; D^{(1)}_{01}(\xi)\;,\nonumber\\
\Sigma_{+3}(\xi)&=& \sqrt{{2\pi^2}/{3}}\;D^{(1)}_{10}(\xi)\;,\nonumber\\
 &\vdots& \nonumber\\
\Sigma_{--}(\xi)&=&\sqrt{{2\pi^2}/{3}}\; D^{(1)}_{-1-1}(\xi)=(\xi_1-i\xi_2)^2 \;.
\end{eqnarray}
This means there are another set of expectation values
\begin{equation}
\langle\langle s:p',q'\vert\Sigma_{ab}\vert s+1:p,q\rangle\rangle \;,
\end{equation}
and their complex conjugates.

For this new class, the analog of Adkins-Nappi-Witten identity are
\begin{equation}\label{ss+1}
\langle\langle s:p',q'\vert\Sigma_{ab}\vert s+1:p,q\rangle\rangle
=-C_1(s)\,\left[{{\cal U}(s:p',q')^\dagger}{\cal U}(s+1:p,q)_{ab}\right] \;,
\end{equation}
with
\begin{eqnarray}
{\cal U}(s+1:p,q)_{ab}&\equiv &(\widehat{\sigma_2\sigma_a})(\widehat{\tau_2\tau_b})
{\cal U}(s+1:p,q)\;,\\
\left((\widehat{\sigma_2\sigma_a})(\widehat{\tau_2\tau_b})
{\cal U}(s+1:p,q)\right)_{\alpha_1\cdots \alpha_{2s}}^{\epsilon_1\cdots \epsilon_{2s}}
&\equiv&
(\sigma_2\sigma_a)^{\beta\beta'}(\tau_2\tau_b)_{\epsilon\epsilon'}
{\cal U}(s+1:p,q)_{\beta\beta' \alpha_1\cdots \alpha_{2s}}^{\epsilon\epsilon'
\epsilon_1\cdots \epsilon_{2s}}\;,\nonumber
\end{eqnarray}
defining the contracting action that reduces the spin (isospin) by one.

To show this identity and determine $C_1(s)$, it again suffices to
compute the case of $p=q=s+1$. Under this restriction, the angular
momentum summation rules tell us that the one and only non-vanishing
matrix element on the left hand side is
\begin{equation}
\langle\langle s:p'=s,q'=s\vert\Sigma_{--}\vert s+1:s+1,s+1\rangle\rangle=
\int_{S^3} (\xi_1-i\xi_2)^2(D^{(s)}_{ss})^*  D^{(s+1)}_{s+1,s+1} \;,
\end{equation}
while $p',q' <s$ producing null results.  Likewise, the right hand side also
vanishes except for
\begin{equation}
-C_1(s)\left[{{\cal U}(s:s,s)^\dagger}\left(
(\widehat{\sigma_2\sigma_-})(\widehat{\tau_2\tau_-}){\cal U}(s+1:s+1,s+1)\right)\right]
=2\times C_1(s) \;,
\end{equation}
since, for any $s$, the only nonvanishing component of ${\cal U}(s:s,s)$ is
\begin{equation}
{\cal U}(s:s,s)^{11\cdots 1}_{11\cdots 1}=1 \;.
\end{equation}
Therefore, the identity in Eq.~(\ref{ss+1}) holds for $p=q=s+1$ with
\begin{eqnarray}
C_1(s)
=\frac12\int_{S^3} (x_1-ix_2)^2(D^{(s)}_{ss})^*  D^{(s+1)}_{s+1,s+1}
=\frac12\sqrt{\frac{2s+1}{2s+3}} \;.
\end{eqnarray}
Similarly with the case of the same spins, the rest of the identity
would follow immediately if the contracting actions by $\tau_2\tau_m$ and
$\sigma_2\sigma_k$ on ${\cal U}$, denoted above as
$\widehat{\sigma_2\sigma_l}$, $\widehat{\tau_2\tau_l}$,
themselves obeys
\begin{equation}
[J_a, \widehat{\sigma_2\sigma_b}]=i\epsilon_{abc}\widehat{\sigma_2\sigma_c}\;,
\end{equation}\begin{equation}
[I_a, \widehat{\tau_2\tau_b}]=i\epsilon_{abc}\widehat{\tau_2\tau_c}\;,
\end{equation}
when acting in the space  of all possible ${\cal U}$'s.

\subsection{Strength of the Magnetic Couplings at Origin}

These generalized identities show that the couplings suggested in the previous section
are indeed exactly the right ones demanded by the instanton origin of the baryons.
They also fix the coefficient functions $h_s$'s and $k_s$'s at origin $w=0$
unambiguously. This was done for the case of $h_{1/2}$ in \cite{Hong:2007ay}, where it
was found to be\footnote{For the proper normalization of the spinors and the coupling,
it is important to recall that the convention for two-component
spinors in this paper is different from that of \cite{Hong:2007ay} where the four-component spinor
was written in terms of two related $\gamma^5$ eigenspinors. Here we are using
two-component spinors which are $\gamma^0$ eigenspinors in the rest frame.}
\begin{equation}
h_{1/2}(w=0)=\frac{2\pi^2}{3}\frac{\rho^2}{e^2(0)} \;.
\end{equation}
If we were considering the instanton soliton in $R^4$,
this coupling would be a constant.

This is straightforwardly generalized to other $h_s$ and $k_s$
 as follows,
\begin{eqnarray}
h_s(0)=2\pi^2\frac{s}{s+1}\,\frac{\rho^2}{e^2(0)},\qquad
k_s(0)=\pi^2\sqrt{\frac{2s+1}{2s+3}}\,\frac{\rho^2}{e^2(0)} \;,
\end{eqnarray}
where the factor $1/3$ in $h_{1/2}$ is replaced by $s^2C_0(s)$ and
by $C_1(s)$, respectively.

The ratio $k_{1/2}(0)/h_{1/2}(0)$
was implicit in Adkins-Nappi-Witten's consideration of
two amplitudes, $\pi N\Delta$ and $\pi N\,N$. The ratio
of the two amplitudes is directly related to the above ratio,
up to normalization issues in terms of defining the amplitudes.
 We took care to verify that the two ratios give the same
physics, which provides an independent check of our computation
of the couplings.

However, the actual geometry is $R^3\times I$ up
to a nontrivial conformal factor as a function of $w$ and this would in general imply
that $h_s$ and $k_s$ are functions of $w$. Due to the fact that a stationary solution
is possible only when the soliton is located at $w=0$, we can determine these
coefficient functions at $w=0$ at best.
In next section, we will finally come to the four-dimensional
physics, and see how these couplings generate cubic and
quartic couplings between baryons and meson in four dimensions.
The fact that these coefficient
functions are not well-determined away from the origin, in general, poses
a systematic difficulty in predicting couplings to excited mesons beyond
those associated with large $N_c$ and large $\lambda$ nature of this model.
For low lying mesons, however, errors  due to this are relatively well controlled.

\section{Baryons Interacting with Mesons}

So far we considered an effective field theory for the instanton soliton of
a fixed size on $R^{4+1}$, using the approximate $SO(4,1)$ symmetry.
This effective field theory is not yet that
of the four dimensional baryons in two aspects. First, even though the
soliton at origin $(w=0)$ sees the approximate $SO(4,1)$ Lorentz symmetry,
a quantum of the spinor fields will see strong breaking of this away from $w=0$.
Second, we must reduce the effective field theory to four dimensions in order to identify the
spinor fields  with
baryons of QCD. In this section, we will incorporate these two issues
and produce a bona fide effective action for QCD baryons.

\subsection{Broken $SO(4,1)$ Symmetry and a Classical Potential for 5D Theory}

The leading effect of having a nontrivial background geometry (conformally $R^{3+1}\times I$)
is that the instanton soliton's mass varies with the position along the holographic
direction. The leading mass comes from
\begin{equation}
\int_{R^3\times I}\frac{1}{8\pi^2 e(w)^2}\,\tr F\wedge F\;,
\end{equation}
and, due to the position-dependence of $1/e(w)^2$, the soliton prefers to sit
near $w=0$. If we have a relativistic formulation, this could be
naturally incorporated into a position dependent mass term.
For nonrelativistic formulation, the mass
shows up as denominator of the quadratic spatial gradient term. Making this
parameter position dependent does not seem to yield the right energetics.

As a toy model
consider a spin 1/2 Dirac field with a position dependent mass
\begin{equation}
-i\bar \Psi \partial_M\gamma^M \Psi + i m(x)\bar\Psi\Psi \;,
\end{equation}
with
\begin{equation}
m(x)=m(s)+V(x),\qquad m(s)>0\; \hbox{and}\; V(x)\ge 0 \;.
\end{equation}
If we are interested in low momentum and low energy behavior of this field,
we may as well treat $V$ as a perturbation. The on-shell condition is then,
\begin{equation}
\left(\begin{array}{cc}
E & i\sigma^m p_m \\
i\bar\sigma^m p_m &-E
\end{array}\right)\Psi
=m(s)\Psi\;,
\end{equation}
with $\sigma^i=\bar\sigma^i$ are the usual Pauli matrices, and $\sigma_4=i=-\bar\sigma_4$.
Using the particle state, for which $E\simeq m(s)+O(p^2)$, and defining the two-component
nonrelativistic spinors
as
\begin{equation}
\Psi
=e^{-im(s) t}
\left(\begin{array}{c}
{\cal U}\\
{\cal V}
\end{array}\right) \;,
\end{equation}
the above relativistic
action reduces to a non-relativistic one as
\begin{equation}
i {\cal U}^\dagger \partial_0{\cal U}
+ \frac{1}{2m(s)} {\cal U}^\dagger \partial_m\partial^m{\cal U}
-V(x){\cal U}^\dagger {\cal U} \;.
\end{equation}
This case of spin 1/2 instructs us, then, to incorporate the
effect of position-dependence of $1/e(w)^2$ as a bilinear of ${\cal U}$
with the coefficient function,
\begin{equation}
V(x)=V(w)=\left(\frac{4\pi^2}{e(w)^2}-\frac{4\pi^2}{e(0)^2}\right)
\end{equation}
in addition to the standard kinetic terms we have. For the baryons, this
acts as a potential in the resulting Schroedinger equation for ${\cal U}$,
which pulls the particles toward the origin $w=0$.

The right thing to do for baryons of any isospin, therefore, is to add
such a potential term to the action, so that the total action is
\begin{equation}
S_{5D}=S_0'+S_I^{magnetic}+S_I^{electric} \;,
\end{equation}
with
\begin{equation}
S_0'=\int dt d^4x\;\left[\sum_s \left( i{\cal U}_s^{\dagger}\frac{\partial}{\partial t}{\cal U}_s
+\sum_{m=1}^4 \frac{1}{2m(s)}{\cal U}_s^{\dagger}(\partial_m-i{\cal A}_m)^2 {\cal U}_s -
V(w){\cal U}_s^\dagger{\cal U}_s\right)\right] \;,
\end{equation}
whereas $S_I^{magnetic}$ is the same interaction piece as in Eq.~(\ref{interaction})
and $S_I^{electric}$ its electric counterpart. For most applications below, we won't
need explicit form of $S_I^{electric}$ since its form and size will be related to
$S_I^{magnetic}$ via Lorentz invariance.

\subsection{Baryon-Meson Couplings from the Dimensional Reduction}

Let's first consider the interactions between mesons and spin
${1}/{2}$ baryons, namely nucleons. In Ref.~\cite{Hong:2007ay} the dimensional reduction
was done for the relativistic theory. Here, we carry out the dimensional
reduction
of non-relativistic theory in 5-dimensions, and we will briefly
compare the two approaches before proceeding to the higher spin cases.
For spin half case, the baryonic wave functions are written as two
component spinors satisfying Schrodinger equation.
Specifically it satisfies
\begin{equation}
i\frac{\partial}{\partial
  t}{\cal U}_{1/2}=-\frac{1}{2m(s)}\left(\partial_i\partial_i+\left(\frac{\partial}{\partial
  w}\right)^2\right){\cal U}_{1/2}+V(w){\cal U}_{1/2} \;.
\end{equation}
If we write it as a product of four-dimensional wave function and the
one dimensional wave function
\begin{equation}
{\cal U}_{1/2}={B}(t, x_i)_{1/2}f(w)e^{-iE_n t} \;,
\end{equation}
where $f(w)$ satisfies the one-dimensional potential problem with
energy eigenvalues $E_n$
\begin{equation}
-\frac{1}{2m(1/2)}\left(\frac{d}{dw}\right)^2f(w)+V(w)f(w)=E_n f(w) \;,
\end{equation}
then $B_{1/2}$ satisfies the free Schrodinger equation
\begin{equation}
i\frac{\partial}{\partial
  t}B_{1/2}=-\frac{1}{2m(1/2)}\partial_i\partial_i B_{1/2} \;,
\end{equation}
where $i=1 \cdots 3$. We are interested in the lowest lying
baryon for each isospin sector, so we will take the smallest
eigenvalue $E_0 $ and its associated ground state wavefunction
$f_0(w)$.

Let us carry out the dimensional reduction of  the following
action describing the interactions between  the spin $\frac{1}{2}$ baryons
and mesons
\begin{eqnarray}
S_{5D}
&=&\int dt\, d^3x \,dw\nonumber\\
&&
  i{\cal U}_{1/2}^{\dagger}(\frac{\partial}{\partial t}-i{\cal A}_0){\cal U}_{1/2}
+\sum_{m=1}^4 \frac{1}{2m(1/2)}{\cal U}_{1/2}^{\dagger}(\partial_m-i{\cal
  A}_m)^2 {\cal U}_{1/2}
-V(w){\cal U}_{1/2}^\dagger{\cal U}_{1/2}   \nonumber \\
 & & -h_{1/2}(w)\left( {\cal U}_{1/2}^{\dagger} \epsilon_{ijk}\sigma_k
 F_{ij} {\cal U}_{1/2}+2{\cal U}_{1/2}^{\dagger} i\sigma_i
 F_{4i} {\cal U}_{1/2} \right)\nonumber \\
 &&-\frac{1}{4e^2(w)}tr {\cal F}_{mn}{\cal F}^{mn} \;,\label{intspin1/2}
\end{eqnarray}
with $F_{ij}=\sum_{a=1}^3 \frac{1}{2}F^{a}_{ij} \tau^a$. Note that the minimal
couplings have $U(2)$ gauge fields while the magnetic
couplings of eq.~(\ref{intspin1/2}) have $SU(2)$ gauge fields, which
source $SU(2)$ instanton fields.
Recall that ${\cal A}_{\mu}(x, w)$ is expanded
in terms of mesons as \cite{sakai-sugimoto}
\begin{equation}
{\cal A}_{\mu}(x, w)= i\alpha_{\mu}(x) \psi_0(w)+i\beta_{\mu}(x)
+\Sigma_{n\ge 1}a_{\mu}^{(n)}(x)\psi_{(n)}(w)\;,
\end{equation}
% =  \sum_{a=1}^3 \left( i\alpha_{\mu}^a(x) \psi_0(w)
%   \frac{\tau^a}{2}
%+i\beta_{\mu}^a(x)\frac{\tau^a}{2}
%+\Sigma_{n\ge 1}a_{\mu}^{a \,(n)}(x)\psi_{(n)}(w) \frac{\tau^a}{2} \right)
with the gauge choice ${\cal A}_w=0$. For the $SU(2)$ part, $A_\mu$,
the first two terms may be expanded in terms of the pion and spin 1
mesons as
%\begin{equation}
%\xi(t,x)=e^{i{\pi(t,x)}/{f_{\pi}}}
%\end{equation}
%$\pi$ is the pion field and $Expanding in terms of the pion field, we
%have
\begin{eqnarray}
A_{\mu}%&=&i\alpha_{\mu}\psi_{0}(w)+i\beta_{\mu}+\cdots  \nonumber \\
% &=& i\{\xi^{-1},
% \partial_{\mu}\xi\}\psi_{0}(w)+\frac{i}{2}[\xi^{-1},
% \partial_{\mu}\xi]+\cdots \nonumber \\
&=& -\frac{2}{f_{\pi}}\partial_{\mu}\pi
 \psi_{0}(w)+\frac{i}{2f_{\pi}^2}[\pi, \partial_{\mu}\pi]+\cdots\;,
\end{eqnarray}
where $f_{\pi}^2=(g^2_{YM}N_c)N_cM^2_{KK}/{54 \pi^4}$
is the pion decay constant.
We also need to separate $U(1)$ part of the vector/axial-vector
mesons as well. Regardless of parity, let us write
\begin{equation}
a_\mu^{(n)}=\left(\begin{array}{cc}N_c/2 &0 \\ 0&N_c/2\end{array}\right)
\omega^{(n)}_\mu+v_\mu^{(n)} \;,
\end{equation}
so the $w$'s are the isosinglets and $v$'s isotriplets.

In the large $\lambda N_c$ limit,  the baryon wave function
is sharply peaked around $w=0$. When we integrate over $w$-direction,
$|f_0(w)|^2 $ may be approximated as a delta function, e.g.,
$\int dw |f_0(w)|^2 \psi_{(n)}(w)=\psi_{(n)}(0)$.
One of the consequence is that the
dimensional reduction of $A_i^2$ can be written as the product of that
of $A_i$, i.e.,
\begin{eqnarray}
& & \int dt\, d^3 x \,dw \sum_{i=1}^3{\cal U}_{1/2}^{\dagger}({\calA}_i)^2 {\cal U}_{1/2}\nonumber \\
&= & \int dt\, d^3 x B_{1/2}^{\dagger}
\left(\alpha_i(t,x)\psi_0(0)+\beta_i(t,x)-i\Sigma_{n\ge
  1}a^{(n)}_i(t,x)\psi_{(n)}(0)\right)^2B_{1/2} \nonumber \\
&=& \int dt\, d^3 x B_{1/2}^{\dagger}
\left(\beta_i(t,x)
-i\Sigma_{n\ge 0}a_i^{(2n+1)}(t,x)\psi_{(2n+1)}(0)\right)^2B_{1/2} \;,
\end{eqnarray}
where we use that $\psi_{(2n)}(w),\psi_{(2n+1)}(w) $ are an odd and
even function of $w$ respectively.

The dimensional reduction of the minimal coupling produce
vector-like couplings as those of $V_{\mu}$
\begin{equation}
S_{minimal}=\int dt\, d^3x
 \left( iB_{1/2}^{\dagger}(\frac{\partial}{\partial t}-iV_0)B_{1/2}
+\sum_{i=1}^3 \frac{1}{2m(1/2)}B_{1/2}^{\dagger}(\partial_i-iV_i)^2 B_{1/2}\right)\;.,
\end{equation}
where $V_\mu$ collects all the vector mesons (as opposed to axial vector mesons)
in their hidden local gauge symmetry \cite{bandoetal,sakai-sugimoto} form
\begin{eqnarray}
V_{\mu}\equiv i\beta_{\mu}(t,x)
+\Sigma_{n \ge 0}a_{\mu}^{(2n+1)}(t,x)\psi_{(2n+1)}(0),  \qquad \mu=0
\cdots 3 \;.
\end{eqnarray}
The coupling constants for couplings to $\beta_{\mu},\omega_{\mu}^{(2n+1)},
v_{\mu}^{(2n+1)} $ are
 1, $N_c\psi_{(2n+1)}(0), \psi_{(2n+1)}(0)$ respectively .
Similarly one can derive the interaction terms coming from five-dimensional magnetic
couplings, which generates the leading couplings
to isotriplet axial vector mesons, $v^{(2n)}$, as well as  derivative couplings
to isotriplet vector mesons, $v^{(2n+1)}$,
\begin{eqnarray}
S_{axial}=&-&\int dt\, d^3x \nonumber \\
&& h_{1/2}(0) 2B_{1/2}^{\dagger} \sigma_i \left(\frac{4i}{\pi}\alpha_i
+\sum_{n\ge 1}v_i^{(2n)}\psi'_{(2n)}(0)\right) B_{1/2} \\
& &+h_{1/2}(0) B_{1/2}^{\dagger}\epsilon_{ijk}\sigma_k
\sum_{n\ge 0}(\partial_iv_j^{(2n+1)}-\partial_jv_i^{(2n+1)})B_{1/2} \;,\nonumber
\end{eqnarray}
where
$\psi'={d\psi}/{dw}$ with $\psi_0'(0)={4}/{\pi}$.
The axial coupling strength is
\begin{equation}
h_{1/2}(0)= \frac{2\pi^2}{3}\frac{\rho^2}{e^2(0)}=\sqrt{\frac{1}{30}}\frac{N_c}{M_{KK}}\;.
\end{equation}
One can see that the interaction terms for $\alpha_{\mu},
v_{\mu}^{(2n)}$ arise only from the 5-dimensional magnetic couplings,
which is observed in relativistic case in the large $N_c$ limit.

%\begin{eqnarray}
%\rho^2 &=& \frac{\sqrt{{2\cdot
%      3^7\cdot\pi^2}/{5}}}{M^2_{KK}(g^2_{YM}N_c)} \nonumber \\
%\frac{1}{e^2(0)}&=&\frac{g^2_{YM}N_c}{108\pi^3}N_cM_{KK},
%\end{eqnarray}
 %we have

It is straightforward to generalize to baryons with general spins.
All of the wavefunctions are sharply peaked around $w$ and
the overlap integrals in $w$ direction act as delta function, which
is true for the overlap integral for wavefunctions of different
spins. The minimal terms are given by
\begin{equation}
S_{minimal}=\int dt\, d^3x
 \sum_s \left( iB_{s}^{\dagger}(\frac{\partial}{\partial t}-iV_0)B_{s}
+\sum_{i=1}^3 \frac{1}{2m_{s}}B_{s}^{\dagger}(\partial_i-iV_i)^2 B_{s}\right)\;,
\end{equation}
where $U_s=U_s(t,x)$ denotes the four-dimensional wave function of the
baryon with spin $s$ from now on.
With $v^{(n)}_{\mu}\equiv \sum_{a=1}^3 v^{(n)\,\,a}_{\mu}\tau^a/2$,
the contribution from magnetic terms (\ref{magneticterm}) are given by
\begin{eqnarray}
&&S_{axial}=\int  dt\, d^3x\nonumber\\
&&-\sum_s \frac12 h_s(0) \sum_{n \ge 0}(\partial_i
v^{(2n+1)}_j-\partial_j v^{(2n+1)}_i)^a
\epsilon_{ijk} (B^{\epsilon'\epsilon_2 \cdots
  \epsilon_{2s}}_{s;\ \, \beta' \alpha_2 \cdots \alpha_{2s}})^*
  \sigma_k^{\beta' \beta} \tau^{\epsilon'\epsilon}_a
B^{\epsilon\epsilon_2 \cdots
  \epsilon_{2s}}_{s;\ \, \beta \alpha_2 \cdots \alpha_{2s}}  \nonumber \\
& &-\sum_s h_s(0) (\frac{4i}{\pi}\alpha_i^a+\sum_{n\ge
  1}v_i^{(2n)\, a}\psi'_{(2n)}(0) )
\,\, (B^{\epsilon'\epsilon_2 \cdots
  \epsilon_{2s}}_{s;\ \, \beta' \alpha_2 \cdots \alpha_{2s}})^*
  \sigma_i^{\beta' \beta} \tau^{\epsilon'\epsilon}_a
B^{\epsilon\epsilon_2 \cdots
  \epsilon_{2s}}_{s;\ \, \beta \alpha_2 \cdots \alpha_{2s}}  \nonumber \\
& & -\sum_s\frac12 k_s(0) \sum_{n\ge 0} (\partial_i
v^{(2n+1)}_j-\partial_j v^{(2n+1)}_i)^a \epsilon_{ijk} (B^{\epsilon_1\epsilon_2 \cdots
  \epsilon_{2s}}_{s;\ \, \alpha_1\alpha_2 \cdots \alpha_{2s}})^*
  (\sigma_2\sigma_k)^{\beta \beta'} (\tau_2\tau_a)^{\epsilon\epsilon'}
B^{\epsilon\epsilon'\epsilon_1 \cdots
  \epsilon_{2s}}_{s+1;\ \, \beta\beta' \alpha_1 \cdots \alpha_{2s}}
  \nonumber \\
& & -\sum_s k_s(0)  (\frac{4i}{\pi}\alpha_i^a+\sum_{n\ge
  1}v_i^{(2n)\, a}\psi'_{(2n)}(0) )  (B^{\epsilon_1\epsilon_2 \cdots
  \epsilon_{2s}}_{s;\ \, \alpha_1\alpha_2 \cdots \alpha_{2s}})^*
  (\sigma_2\sigma_i)^{\beta \beta'} (\tau_2\tau_a)^{\epsilon\epsilon'}
B^{\epsilon\epsilon'\epsilon_1 \cdots
  \epsilon_{2s}}_{s+1;\ \, \beta\beta' \alpha_1 \cdots \alpha_{2s}}\;.\nonumber \\
  &&
\end{eqnarray}
where
\begin{eqnarray}
h_s(0)&=& 2\pi^2\frac{s}{s+1}\frac{\rho^2}{e^2(0)}
=\frac{\sqrt{6}}{2\sqrt{5}}\frac{s}{s+1}\frac{N_c}{M_{KK}} \;,\nonumber \\
k_s(0)&=& \pi^2\sqrt{\frac{2s+1}{2s+3}}\frac{\rho^2}{e^2(0)}
=\frac{\sqrt{6}}{4\sqrt{5}}\sqrt{\frac{2s+1}{2s+3}}\frac{N_c}{M_{KK}}\;.
\end{eqnarray}

\subsection{Baryon-Pion Interactions}

These two sets of interaction terms contain couplings to the
pion field $\pi(x)$ from
\begin{equation}
A_{\mu}^{a}=-\frac{2}{f_{\pi}}\partial_{\mu}\pi^a\psi_0(0)
-\frac{1}{2f_{\pi}^2}\varepsilon^{abc}\pi^b\partial_{\mu}\pi^c+\cdots.
\end{equation}
{}From  $S_{minimal}$, we find
\begin{eqnarray}
&&\sum_s B_s^{\dagger}(-\frac{1}{2f_{\pi}^2}\varepsilon^{abc}\pi^b
  \partial_0 \pi^c\frac{\tau^a}{2}) B_{s}  \nonumber \\
& &+ \frac{1}{2m(s)}\sum_{i=1}^3  B_{s}^{\dagger}
(\frac{i}{f_{\pi}^2}\varepsilon^{abc}\pi^b\partial_i \pi^c\partial_i
+\frac{i}{2f_{\pi}^2}\varepsilon^{abc}\pi^b\partial_i \partial_i\pi^c
  ) \frac{\tau^a}{2} B_{s}  \nonumber \\
&&-
  \frac{1}{4f_{\pi}^4}B_{s}^\dagger(\epsilon_{abc}\pi^b\partial_i\pi^c\frac{\tau^a}{2}
\epsilon_{def}\pi^e\partial_i\pi^f\frac{\tau^d}{2})B_{s} +\cdots\;,
\end{eqnarray}
up to terms higher order in $1/f_\pi$,
where the gauge generators $\tau_a$'s act only on the first
gauge doublet index of $B_s$'s, and, from $S_{axial}$
\begin{eqnarray}
&&\sum_s h_s(0) \frac{8}{\pi f_{\pi}}\partial_i\pi^a
\,\, (B^{\epsilon'\epsilon_2 \cdots
  \epsilon_{2s}}_{s;\ \, \beta' \alpha_2 \cdots \alpha_{2s}})^*
  \sigma_i^{\beta' \beta} \tau^{\epsilon'\epsilon}_a
B^{\epsilon\epsilon_2 \cdots
  \epsilon_{2s}}_{s;\ \, \beta \alpha_2 \cdots \alpha_{2s}}  \nonumber \\
&&+\sum_s  h_s(0)
\frac{1}{2f_{\pi}^2}\varepsilon^{abc}\partial_i\pi^b\partial_j\pi^c
\epsilon_{ijk} (B^{\epsilon'\epsilon_2 \cdots
  \epsilon_{2s}}_{s;\ \, \beta' \alpha_2 \cdots \alpha_{2s}})^*
  \sigma_k^{\beta' \beta} \tau^{\epsilon'\epsilon}_a
B^{\epsilon\epsilon_2 \cdots
  \epsilon_{2s}}_{s;\ \, \beta \alpha_2 \cdots \alpha_{2s}}  \nonumber \\
  & &+ \sum_s k_s(0) \frac{8}{\pi f_{\pi}}\partial_i\pi^a
   (B^{\epsilon_1\epsilon_2 \cdots
  \epsilon_{2s}}_{s;\ \, \alpha_1\alpha_2 \cdots \alpha_{2s}})^*
  (\sigma_2\sigma_i)^{\beta \beta'} (\tau_2\tau_a)^{\epsilon\epsilon'}
B^{\epsilon\epsilon'\epsilon_1 \cdots
  \epsilon_{2s}}_{s+1;\ \, \beta\beta' \alpha_1 \cdots \alpha_{2s}}  \nonumber\\
&&+  \sum_s k_s(0) \frac{1}{2f_{\pi}^2}\varepsilon^{abc}\partial_i\pi^b\partial_j\pi^c
\epsilon_{ijk} (B^{\epsilon_1\epsilon_2 \cdots
  \epsilon_{2s}}_{s;\ \, \alpha_1\alpha_2 \cdots \alpha_{2s}})^*
  (\sigma_2\sigma_k)^{\beta \beta'} (\tau_2\tau_a)^{\epsilon\epsilon'}
B^{\epsilon\epsilon'\epsilon_1 \cdots
  \epsilon_{2s}}_{s+1;\ \, \beta\beta' \alpha_1 \cdots \alpha_{2s}}
  \nonumber \\
  &&+\cdots \;,
\end{eqnarray}
again up to terms higher order in $1/f_\pi$.
These are generalization of pion-nucleon couplings, which
altogether may be  written compactly as
\begin{eqnarray}
&& h_{1/2}(0) \frac{16}{\pi f_{\pi}}
B_{1/2}^{\dagger}\sigma_i \partial_i\pi B_{1/2}  - h_{1/2}(0)B_{1/2}^{\dagger}\epsilon_{ijk}\sigma_k
\frac{i}{2 f_{\pi}^2}[\partial_i \pi, \partial_j \pi]B_{1/2} \nonumber \\
&& +\frac{i}{2f_{\pi}^2}B_{1/2}^{\dagger}[\pi,
\partial_0\pi]B_{1/2}+\frac{1}{2m(1/2)}
\frac{1}{2f_{\pi}^2}B_{1/2}^{\dagger}([\pi, \partial_i\partial_i \pi]+2[\pi, \partial_i
\pi]\partial_i)B_{1/2} \nonumber  \\
&&+ \frac{1}{2m_0}B_{1/2}^{\dagger}(\frac{1}{4f_{\pi}^4}
[\pi,\partial_i\pi][\pi, \partial_i\pi])B_{1/2} +\cdots\;,
\end{eqnarray}
similarly in the $1/f_\pi$ expansion.

\subsection{A Comment on Subleading Corrections and Relativistic Formulation}

One major difference between the relativistic and the nonrelativistic
approaches is a loss, or ambiguity, of subleading terms. A good
illustration of this is the leading axial coupling to the pion.
In the large $N_c$ limit, the magnetic coupling gives the dominant
contribution scaling linearly with $N_c$. The relativistic
kinetic term, however, also contribute $O(1)$ term inversely
proportional to the mass of the baryon. The mechanism
behind the latter is precisely the same as how one obtains $g=2$
nonanomalous gyromagnetic ratio from the minimal coupling
of a Dirac fermion to electromagnetic gauge field.

Once we abandon
the relativistic formulation, therefore, such terms can only
be included in the interaction terms somewhat arbitrarily. Just as
one cannot predict $g=2$ from Schroedinger equation, we cannot
compute the subleading term to pion-baryon coupling due to
the minimal coupling. This problem is not confined to the
pion coupling and is applicable to all terms we are considering.
Obviously this does not affect our leading contributions, but
it also tells us that finding a fully relativistic form is
essential for improving the present result to next order.
We should note that this sort of problem also manifest itself
in computation of four-dimensional mass of the baryons.

\section{Summary}

In this paper, we generalize the derivation of the interactions
between mesons and nucleons carried out in Ref.~\cite{Hong:2007kx,Hong:2007ay}
to the interactions between mesons and baryons of arbitrary
half-integer spins for the two-flavor case ($N_F=2$).
Following the approach given in Ref.~\cite{Hong:2007kx}, we resort
to the instantonic origin of the baryon fields, and produced general
prescriptions and formulae that determine the strength of each
interaction term in the large $N_c$ and large $\lambda$ limit.
For the nucleon case (isospin 1/2), this program
has produced a relativistic action and a rich phenomenology
\cite{Hong:2007kx,Hong:2007ay,Hong:2007dq}, improving
the Skyrme model computations by Adkins, Nappi and Witten \cite{ANW} substantially.

Baryons are realized holographically as small instanton solitons
in five dimensions with a Coulombic hair,
whose quantization gives rise to baryons of (half-)integer spins.
The corresponding on-shell field content may be realized as fermionic
fields with
symmetric spinor indices under the little group as well as the same
number of symmetric
isospin indices under the flavor group.
However  it's not clear how to write down the  relativistic
action since the relativistic version of such multi-spinor fermion
is not known. The main difficulty is in finding a relativistic
formulation where the appropriate constraints may be built in
at the level of action. Due to this
technical difficulty, we chose to consider the
non-relativistic limit for the baryons instead. This limit is sufficient, as
it turned out, if we look only for the leading large $N_c$ results.

Since the spin fields arise from the quantization of the instanton,
 the interaction terms between the holographic baryons and the
 five-dimensional flavor gauge fields should be compatible with the
 semiclassical instantonic
 configuration. Out of this consideration, a single term, called
 the magnetic term, together with
 the usual minimal coupling essentially determines all the interactions
between the mesons and baryons upon the dimensional reduction of
the five dimensional nonrelativistic actions down to four dimensions
because the dimensional reduction of $U(2)$ gauge field give rise
to towers of mesons including pion fields.
In particular, when restricted to the sector of nucleons,
the nonrelativistic approach adopted  here reproduces
the same results as in Ref.~\cite{Hong:2007ay} derived from the
relativistic case in the large $N_c$ limit .

For subleading corrections in $1/N_c$, which would be relevant
for  $N_c=3$ case as appropriate for real QCD, some ambiguities
remain in part because we had to use nonrelativistic
formulation and also in part because of other more fundamental reasons.
These include other $1/N_c$ corrections (notably the one due to
quenching and also due to the inherent limitations present in
any AdS/QCD models) which are not well understood either.

We hope that this work will provide the starting point for comparing
with the experimental data or other field theoretical computations on
the interactions between mesons and baryons.
Finally all of the baryons we consider have just $SU(2)$ isospin
symmetry. It would be interesting to extend the current work
to $SU(3)$ case, which would include strange baryons.

\vskip 1.5cm

\centerline{\bf\large Acknowledgement}
\vskip 5mm\noindent
We thank Ho-Ung Yee for a usueful comment on the manuscript.
This work  is supported in part by the Science Research Center Program
of KOSEF through the Center for Quantum Space-Time (CQUeST) of
Sogang-University with the grant number R11-2005-021 (J.P.,P.Y.),
by the Postech BSRI research fund 2007 (J.P.), and by the Korea
Research Foundation Grant funded by the Korean Government
(MOEHRD, Basic Research Promotion Fund, KRF-2007-314-C00052) (P.Y.).
Both authors are also supported by the Stanford Institute for
Theoretical Physics (SITP Quantum Gravity visitor fund).


\begin{thebibliography}{99}




\bibitem{Maldacena:1997re}
  J.~M.~Maldacena,
  ``The large N limit of superconformal field theories and supergravity,''
  Adv.\ Theor.\ Math.\ Phys.\  {\bf 2}, 231 (1998)
  %[Int.\ J.\ Theor.\ Phys.\  {\bf 38}, 1113 (1999)]\\
 [arXiv:hep-th/9711200];\\
  %%CITATION = HEP-TH 9711200;%%
  %%Cited 1362 times in SPIRES-HEP
%\cite{Gubser:1998bc}
%\bibitem{Gubser:1998bc}
  S.~S.~Gubser, I.~R.~Klebanov and A.~M.~Polyakov,
  ``Gauge theory correlators from non-critical string theory,''
  Phys.\ Lett.\  B {\bf 428}, 105 (1998)
  [arXiv:hep-th/9802109];\\
  %%CITATION = PHLTA,B428,105;%%
  %\cite{Witten:1998qj}
%\bibitem{Witten:1998qj}
  E.~Witten,
  ``Anti-de Sitter space and holography,''
  Adv.\ Theor.\ Math.\ Phys.\  {\bf 2}, 253 (1998)
[arXiv:hep-th/9802150].
  %%CITATION = 00203,2,253;%%

%
\bibitem{Witten:1998zw}
  E.~Witten,
  ``Anti-de Sitter space, thermal phase transition, and confinement in  gauge
  theories,''
  Adv.\ Theor.\ Math.\ Phys.\  {\bf 2}, 505 (1998)
  [arXiv:hep-th/9803131].
  %%CITATION = 00203,2,505;%%

\bibitem{Csaki:1998qr}
  C.~Csaki, H.~Ooguri, Y.~Oz and J.~Terning,
  ``Glueball mass spectrum from supergravity,''
  JHEP {\bf 9901}, 017 (1999)
  [arXiv:hep-th/9806021].
  %%CITATION = JHEPA,9901,017;%%
%\cite{}
\bibitem{Brower:2000rp}
  R.~C.~Brower, S.~D.~Mathur and C.~I.~Tan,
  ``Glueball spectrum for QCD from AdS supergravity duality,''
  Nucl.\ Phys.\  B {\bf 587}, 249 (2000)
  [arXiv:hep-th/0003115].
  %%CITATION = NUPHA,B587,249;%%

\bibitem{sakai-sugimoto}
  T.~Sakai and S.~Sugimoto,
  ``Low energy hadron physics in holographic QCD,''
  Prog.\ Theor.\ Phys.\  {\bf 113}, 843 (2005)
  [arXiv:hep-th/0412141];\\
  %%CITATION = HEP-TH 0412141;%%
%
%\cite{Sakai:2005yt}
%\bibitem{Sakai:2005yt}
  T.~Sakai and S.~Sugimoto,
  ``More on a holographic dual of QCD,'' Prog.\ Theor.\ Phys.\
  {\bf 114}, 1083 (2006)
  [arXiv:hep-th/0507073].
  %%CITATION = HEP-TH 0507073;%%

\bibitem{skyrme} T.H.R. Skyrme, `` A unified field theory of mesons and
baryons," {Nucl. Phys.}\ {\bf 31}, 556 (1962).

\bibitem{Hashimoto:2007ze}
  K.~Hashimoto, C.~I.~Tan and S.~Terashima,
  ``Glueball Decay in Holographic QCD,''
  arXiv:0709.2208 [hep-th].
  %%CITATION = ARXIV:0709.2208;%%

\bibitem{Hong:2007kx}
  D.~K.~Hong, M.~Rho, H.~U.~Yee and P.~Yi,
  ``Chiral dynamics of baryons from string theory,''
  Phys. Rev. D {\bf 76}, 061901, (2007)
  [arXiv:hep-th/0701276].
  %%CITATION = HEP-TH/0701276;%%

\bibitem{Hata:2007mb}
  H.~Hata, T.~Sakai, S.~Sugimoto and S.~Yamato,
  ``Baryons from instantons in holographic QCD,''
  [arXiv:hep-th/0701280].
  %%CITATION = HEP-TH/0701280;%%
%\cite{Hong:2006ta}



%\cite{Hong:2007ay}
\bibitem{Hong:2007ay}
  D.~K.~Hong, M.~Rho, H.~U.~Yee and P.~Yi,
  ``Dynamics of Baryons from String Theory and Vector Dominance,''
  JHEP {\bf 0709}, 063 (2007)
  [arXiv:0705.2632 [hep-th]].
  %%CITATION = JHEPA,0709,063;%%


%
\bibitem{Kim:2008zn}
  K.~Y.~Kim, S.~J.~Sin and I.~Zahed,
  %``Dense and Hot Holographic QCD: Finite Baryonic E Field,''
  arXiv:0803.0318 [hep-th].
  %%CITATION = ARXIV:0803.0318;%%

\bibitem{Bergman:2008sg}
  O.~Bergman, G.~Lifschytz and M.~Lippert,
  %``Response of Holographic QCD to Electric and Magnetic Fields,''
  arXiv:0802.3720 [hep-th].
  %%CITATION = ARXIV:0802.3720;%%

\bibitem{Kim:2007zm}
  K.~Y.~Kim, S.~J.~Sin and I.~Zahed,
  %``The Chiral Model of Sakai-Sugimoto at Finite Baryon Density,''
  JHEP {\bf 0801}, 002 (2008)
  [arXiv:0708.1469 [hep-th]].
  %%CITATION = JHEPA,0801,002;%%

%\cite{Hong:2007dq}
\bibitem{Hong:2007dq}
  D.~K.~Hong, M.~Rho, H.~U.~Yee and P.~Yi,
  ``Nucleon Form Factors and Hidden Symmetry in Holographic QCD,''
  Phys.\ Rev.\  D {\bf 77}, 014030 (2008)
  [arXiv:0710.4615 [hep-ph]].
  %%CITATION = PHRVA,D77,014030;%%


%\cite{Witten:1998xy}
\bibitem{witten-baryon}
  E.~Witten,
  ``Baryons and branes in anti de Sitter space,''
  JHEP {\bf 9807}, 006 (1998)
  [arXiv:hep-th/9805112].
  %%CITATION = JHEPA,9807,006;%%  %\cite{Hong:2007kx}



%
\bibitem{Atiyah:1989dq}
  M.~F.~Atiyah and N.~S.~Manton,
  ``Skyrmions from instantons,''
  Phys.\ Lett.\  B {\bf 222}, 438 (1989).
  %%CITATION = PHLTA,B222,438;%%



\bibitem{vector-skyrmion} T. Fujiwara et al, `` An effective Lagrangian
for pions, $\rho$ mesons and skyrmions," Theor. Phys. {\bf 74},
128 (1985);\\ U.-G. Meissner, N. Kaiser, A. Wirzba and W. Weise,
`` Skyrmions with $\rho$ and $\omega$ mesons as dynamical gauge
bosons," Phys. Rev. Lett. {\bf 57}, 1676 (1986);\\ U.G. Meissner
and I. Zahed, `` Skyrmions in the presence of vector mesons,"
{\prl}\ {\bf 56}, 1035 (1986).

%\cite{Nawa:2007gh}
\bibitem{Nawa:2007gh}
  K.~Nawa, H.~Suganuma and T.~Kojo,
  ``Brane-induced Skyrmions: Baryons in holographic QCD,''
  [arXiv:hep-th/0701007].
  %%CITATION = HEP-TH/0701007;%%
\bibitem{bandoetal}  M. Bando, T. Kugo and K. Yamawaki, ``
Nonlinear realization and hidden local symmetries." Phys. Rept.
{\bf 164}, 217 (1988).

\bibitem{ANW} G.S. Adkins, C.R. Nappi and E. Witten,  `` Static
properties of nucleons in the Skyrme model," Nucl. Phys. {\bf
B228}, 552 (1983).

\bibitem{Witten:1983tx}
  E.~Witten,
  ``Current Algebra, Baryons, And Quark Confinement,''
  Nucl.\ Phys.\  B {\bf 223}, 433 (1983).
  %%CITATION = NUPHA,B223,433;%%
\bibitem{Finkelstein:1968hy}
  D.~Finkelstein and J.~Rubinstein,
  ``Connection between spin, statistics, and kinks,''
  J.\ Math.\ Phys.\  {\bf 9}, 1762 (1968).
  %%CITATION = JMAPA,9,1762;%%





\end{thebibliography}
\end{document}